\documentclass{aa}

\usepackage{epsfig}
\usepackage{epstopdf}
\usepackage{multirow}
\usepackage{caption}
\usepackage[varg]{txfonts}

\def \aap  {A\&A}

\def \apjs  {ApJS}
\def \apjl  {ApJL}

\def \chisq  {\ifmmode  \chi^2   \else  $\chi^2$  \fi}  
\def \spose#1{\hbox  to 0pt{#1\hss}}  
\def \lta{\mathrel{\spose{\lower 3pt\hbox{$\sim$}}\raise  2.0pt\hbox{$<$}}}
\def \gta{\mathrel{\spose{\lower  3pt\hbox{$\sim$}}\raise 2.0pt\hbox{$>$}}}

\def \kms {\ifmmode  \,\rm km\,s^{-1} \else $\,\rm km\,s^{-1}  $ \fi }
\def \kpc {\ifmmode  {\rm~kpc}  \else ${\rm~kpc}$\fi}  
\def \pc {\ifmmode  {\rm~pc}  \else ${\rm~pc}$ \fi  }  
\def \Gyr {\ifmmode  {\rm~Gyr}  \else ${\rm~Gyr}$\fi}
\def \Msun {\ifmmode M_{\odot} \else $M_{\odot}$ \fi} 
\def \Lsun {\ifmmode L_{\odot} \else $L_{\odot}$ \fi} 
\def \Rsun {\ifmmode R_{\odot} \else $R_{\odot}$ \fi} 
\def \Msunpyr {\ifmmode M_{\odot}{\rm~yr}^{-1} \else $M_{\odot}{\rm~yr}^{-1}$ \fi} 
\def \hMsun {\ifmmode h^{-1}\,\rm M_{\odot} \else $h^{-1}\,\rm M_{\odot}$ \fi}

\def \LCDM {\ifmmode \Lambda{\rm CDM} \else $\Lambda{\rm CDM}$ \fi}
\def \sig8 {\ifmmode \sigma_8 \else $\sigma_8$ \fi} 
\def \OmegaM {\ifmmode \Omega_{\rm M} \else $\Omega_{\rm M}$ \fi} 
\def \OmegaL {\ifmmode \Omega_{\rm \Lambda} \else $\Omega_{\rm \Lambda}$\fi} 
\def \Deltavir {\ifmmode \Delta_{\rm vir} \else $\Delta_{\rm vir}$ \fi}
\def \rhocrit {\ifmmode \rho_{\rm crit} \else $\rho_{\rm crit}$ \fi}
\def \rhou {\ifmmode \rho_{\rm u} \else $\rho_{\rm u}$ \fi}
\def \zc {\ifmmode z_{\rm c} \else $z_{\rm c}$ \fi}

\def \rhos {\ifmmode \rho_{\rm s} \else $\rho_{\rm s}$ \fi} 
\def \rs {\ifmmode r_{\rm s} \else $r_{\rm s}$ \fi} 
\def \cvir {\ifmmode c_{\rm vir} \else $c_{\rm vir}$ \fi} 
\def \Rvir {\ifmmode r_{\rm vir} \else $R_{\rm vir}$ \fi}
\def \Vvir {\ifmmode V_{\rm  vir} \else  $V_{\rm vir}$  \fi} 
\def \Mvir {\ifmmode M_{\rm  vir} \else $M_{\rm  vir}$ \fi}  
\def \Nvir {\ifmmode N_{\rm  vir} \else $N_{\rm  vir}$ \fi}  
\def \Jvir {\ifmmode J_{\rm vir} \else $J_{\rm vir}$ \fi} 
\def \Evir {\ifmmode E_{\rm vir} \else $E_{\rm vir}$ \fi} 
\def \vvir {\ifmmode v_{\rm vir} \else $v_{\rm vir}$ \fi} 
\def \lam {\ifmmode \lambda  \else $\lambda$ \fi} 
\def \lamp {\ifmmode \lambda^{\prime} \else $\lambda^{\prime}$  \fi} 
\def \Vmax {\ifmmode V_{\rm  max} \else  $V_{\rm max}$  \fi} 
\def \Mdm {\ifmmode M_{\rm  dm} \else $M_{\rm  dm}$ \fi}

\def \Mgas {\ifmmode M_{\rm gas} \else $M_{\rm gas}$ \fi} 
\def \Mcg {\ifmmode M_{\rm cg} \else $M_{\rm cg}$\fi} 
\def \Mhg {\ifmmode M_{\rm hg} \else $M_{\rm hg}$ \fi} 
\def \Mdisc {\ifmmode M_{\rm disc} \else $M_{\rm disc}$ \fi} 
\def \Md {\ifmmode M_{\rm d} \else $M_{\rm d}$ \fi} 
\def \Mda {\ifmmode M_{\rm d,0\%} \else $M_{\rm d,0\%}$ \fi} 
\def \Mdb {\ifmmode M_{\rm d,20\%} \else $M_{\rm d,20\%}$ \fi} 
\def \Mdc {\ifmmode M_{\rm d,40\%} \else $M_{\rm d,40\%}$ \fi} 
\def \md {\ifmmode m_{\rm d} \else $m_{\rm d}$ \fi} 
\def \Mb {\ifmmode M_{\rm b} \else $M_{\rm b}$ \fi} 
\def \Mbh {\ifmmode M_{\rm b,pri} \else $M_{\rm b,pri}$ \fi} 
\def \Mbs {\ifmmode M_{\rm b,sat} \else $M_{\rm b,sat}$ \fi} 
\def \zo {\ifmmode z_{0} \else $z_{0}$ \fi} 
\def \rd {\ifmmode r_{\rm d} \else $r_{\rm d}$ \fi}
\def \rg {\ifmmode r_{\rm g} \else $r_{\rm g}$ \fi}
\def \rb {\ifmmode r_{\rm b} \else $r_{\rm b}$\fi}
\def \rs {\ifmmode r_{\rm s} \else $r_{\rm s}$\fi}
\def \rc {\ifmmode r_{\rm c} \else $r_{\rm c}$\fi}
\def \rvir {\ifmmode r_{\rm vir} \else $r_{\rm vir}$\fi}
\def \rbh {\ifmmode r_{\rm b,pri} \else $r_{\rm b,pri}$ \fi} 
\def \rbs {\ifmmode r_{\rm b,sat} \else $r_{\rm b,sat}$ \fi} 
\def \zp {\ifmmode z_{\rm phot} \else $z_{\rm phot}$ \fi}
\def \zs {\ifmmode z_{\rm spec} \else $z_{\rm spec}$ \fi}
\def \Lya{\ensuremath{\mathrm{Ly}\alpha\ }}

\begin{document}

\title{Rest-frame ultraviolet spectra of massive galaxies at $z\sim3$: evidence of high-velocity outflows} 

\author{Wouter  Karman\inst{\ref{inst1}}\thanks{karman@astro.rug.nl} \and Karina I. Caputi\inst{\ref{inst1}} \and Scott C. Trager\inst{\ref{inst1}} \and Omar Almaini\inst{\ref{inst2}} \and Michele Cirasuolo\inst{\ref{inst3},\ref{inst4}}}
\institute{ Kapteyn Astronomical Institute, University of Groningen, Postbus 800, 9700 AV Groningen, the Netherlands\label{inst1} 
\and School of Physics and Astronomy, University of Nottingham, University Park, Nottingham NG7 2RD, UK \label{inst2}
\and SUPA, Institute for Astronomy, The University of Edinburgh, Royal Observatory, Edinburgh, EH9 3HJ, UK \label{inst3}
\and UK Astronomy Technology Centre, Royal Observatory, Blackford Hill, Edinburgh, EH9 3HJ, UK \label{inst4}
}

\date{Received ... /
Accepted 17 February 2014}

\label{firstpage}

\abstract{
Galaxy formation models invoke the presence of strong feedback mechanisms that regulate the growth of massive galaxies at high redshifts. Providing observational evidence of these processes is crucial to justify and improve these prescriptions. 
In this paper we aim to (1) confirm spectroscopically the redshifts of a sample of massive galaxies selected with photometric redshifts $\zp>2.5$; (2) investigate the properties of their stellar and interstellar media; 
(3) detect the presence of outflows and measure their velocities.
To achieve this, we analysed deep, high-resolution ($R\approx2000$) FORS2 rest-frame UV spectra for 11 targets. We confirmed that 9 out of 11 have spectroscopic redshifts $\zs>2.5$. We also serendipitously found two mask fillers at redshift $\zs>2.5$, which originally were assigned photometric redshifts $2.0<\zp<2.5$. 
In the four highest quality spectra we derived outflow velocities by fitting the absorption line profiles with models including multiple dynamical components. We found strongly asymmetric, high-ionisation lines, from which we derived outflow velocities ranging between $480$~km s$^{-1}$ and $1518$~km s$^{-1}$. The two highest velocity outflows correspond to galaxies with active galactic nuclei (AGNs). We revised the spectral energy distribution fitting $U$-band through 8~$\mu$m photometry, including the analysis of a power-law component subtraction to identify the possible presence of AGNs. The revised stellar masses of all but one of our targets are $\gta 10^{10} \rm \Msun$, with four having stellar masses $> 5\times 10^{10} \rm \Msun$.  Three galaxies have significant power-law components in their spectral energy distributions, indicating that they host AGNs.
We conclude that massive galaxies are characterised by significantly higher velocity outflows than the typical Lyman-break galaxies at $z\sim3$. The incidence of high-velocity outflows ($\sim40$\% within our sample) is also much higher than among massive galaxies at $z<1$, consistent with the powerful star formation and nuclear activity that most massive galaxies display at $z>2$.
}

\keywords{
Galaxies: high redshift, active, ISM, evolution--
ISM: jets and outflows--
Techniques: spectroscopic
}

\titlerunning{UV spectra of massive galaxies at $z \sim 3$}
\authorrunning{W. Karman et al.}

\setcounter{footnote}{1}

\maketitle

\section{Introduction}
\label{sec:intro}

Understanding the formation and evolution of massive galaxies ($M_\star > 10^{10} M_\odot$) at high redshifts is important for constraining galaxy formation models. Early hierarchical models of galaxy formation  
were unable to explain the large number of massive galaxies found at redshifts $z > 1$ \citep[e.g. ][]{Cimatti2002,Drory2003,Pozzetti2003}. Over the last ten years, the implementation and improvement of feedback recipes in these models has progressively led to a better agreeement with observational results \citep{Somerville2004, Bower2006,Croton2006}. Feedback is now considered to be a crucial component of galaxy formation models. This feedback consists of winds produced by massive stars, supernovae (SN), and galactic nuclei that regulate the molecular gas collapse leading to new star formation. However, our knowledge of feedback effects in galaxies at different redshifts is still sparse,  and as a consequence this effect is loosely implemented in theoretical models. 

Several studies have shown that a significant fraction ($\sim$20-40\%) of the massive galaxies that we see today were already in place and massive by a redshift of $z\sim2$ \citep[e.g.][]{Daddi2005,Caputi2006a}. More recently,  \citet{Caputi2011} determined that this percentage is much lower ($\sim3-5\%$) at $z\sim3$, and decreases even faster at higher redshifts, becoming almost negligible by redshift $z\sim5$. Therefore, the short cosmic time elapsed between redshifts $z\sim 5$ and $z\sim 2$ appears to have been the most efficient epoch for massive galaxy assembly. During this epoch, gaseous inflows and outflows are expected to be strong since they closely regulate star formation. The inflows fuel a large gas supply that enables high star formation rates (SFR) and must therefore have been accompanied by powerful gas outflows that eventually ceased the star formation activity \citep[e.g.][]{Efstathiou2000,Croton2006,Sales2010,Hopkins2012b}. This produced the mostly passive, massive galaxies that we see today. Thus, searching for evidence of gas outflows at $2<z<5$ is important to confirm the validity of this proposed mechanism that regulates massive galaxy growth. 

Massive galaxies have been widely studied up to redshifts $z \sim 2$, covering the period after the main peak of global star formation activity. These studies have investigated the properties of massive galaxies, such as age, stellar composition, dust reddening, and morphology, and also the formation of the red sequence after star-formation quenching \citep[e.g.][]{Franx2003,Saracco2005, Tasca2009,Cirasuolo2010,Cucciati2010}. Moreover, extensive spectroscopic campaigns have provided redshift confirmation for large samples of galaxies up to $z\sim2$  \citep[e.g.][]{Cimatti2002,LeFevre2005,Lilly2007,Kurk2013}. At higher redshifts, the amount of spectroscopic data is much smaller, and is still biased towards optically selected intermediate or low-mass galaxies \citep[e.g.][]{Steidel2003,Vanzella2009}. A systematic spectroscopic study of massive galaxies at $z>2.5$ is still lacking in the literature and is essential to i) confirm the validity of galaxy studies at these redshifts, which are still largely based on photometric redshifts and ii) study in detail the physical conditions of early massive galaxy formation.

The presence of outflows is among the physical properties that have been extensively investigated at low and intermediate redshifts. \citet{Martin2005}, \citet{Rupke2005a,Rupke2005b}, and \citet{Cicone2012} found outflows with velocities up to and over 1500~\kms in ultra-luminous infrared galaxies (ULIRGs) at redshifts up to $z < 0.5$. \citet{Diamond-Stanic2012} found outflow velocities in excess of 1000~\kms  in a sample of obscured compact starburst galaxies at $z\sim0.6$, half of which are luminous infrared galaxies (LIRGs) or ULIRGs. In contrast, \citet{Weiner2009} derived an average outflow velocity of only 300~\kms for a composite spectrum of an opticallyselected sample of galaxies with $\log{M_\star} > 10.45$ at $z \sim 1.4$. \citet{Weiner2009}, and \citet{Bordoloi2013} found increasing outflow velocities for composite spectra with increasing SFRs. \citet{Bradshaw2013} studied outflow velocities as a function of several properties. They confirm increase of velocity with SFR and mass, but also found correlations with the specific star formation rate (SSFR, defined as SSFR$=$SFR/$M_{\star}$). All these studies taken together suggest that the outflow velocities measured for massive galaxies are strongly dependent on their level of activity, as one would expect. 

Ultra-luminous infrared galaxies and luminous infrared galaxies are increasingly rare at $z<1.5$ and $z<1$, respectively, and are not representative of the bulk of massive galaxies at low redshifts, and this is why the high-velocity outflows found in LIRGs/ULIRGs are not indicative of the typical properties of the massive galaxy interstellar media (ISM). At redshifts $z>2$, a substantial fraction of massive galaxies are ULIRGs  \citep[e.g.][]{Caputi2006b,Rodighiero2010,Magnelli2011,Murphy2011}, suggesting that the presence of outflows should be ubiquitous among massive systems.  However, as most spectroscopic studies at $z>2$ mainly targeted Lyman-break galaxies, which mostly have stellar masses $M_{\star}<10^{10} \, \rm M_\odot$, the derived outflow velocities are typically moderate.  For example, \citet{Shapley2003} found outflow velocities in the range 200-600 \kms, \citet{Verhamme2008} reported 150-200~\kms, and \citet{Steidel2010} found maximum outflow velocities of 800~\kms in their spectral samples. Only a few studies focused on more massive, active galaxies presented evidence for larger outflow velocities. \citet{Harrison2012} found high-velocity outflows of up to 1400~\kms in four ULIRGs hosting active galactic nuclei (AGNs) at a median redshift of $z\sim2.3$. \citet{Nesvadba2008} found outflow velocities of $\sim$1000~\kms in three powerful radio galaxies at $2.4\leq z \leq 3.1$. 

In this paper we study medium-resolution ($R\approx2000$) rest-UV spectra of 11 massive galaxies at $z>2.5$. In Sect. \ref{sec:data} we give details of our target selection and spectroscopic data reduction. In Sect. \ref{sec:results} we present the analysis of all our spectra on an individual basis and perform a detailed outflow modelling of our four best quality spectra in Sect. \ref{sec:outlowanalysis}. In this section, we also compare our results with previous studies available in the literature. In Sect. \ref{sec:models} we revise the spectral energy distribution modelling of our targets based on our newly derived spectroscopic redshifts. Finally, in Sect. \ref{sec:discussion} we summarise our findings and present some concluding remarks. Throughout this paper, we adopt a cosmology with $H_0~=~70~$\kms~ Mpc$^{-1}$, $\Omega_M~=~0.3$, and $\Omega_\Lambda~=~0.7$. All magnitudes refer to the AB system, and we use a Salpeter (1955) initial mass function over stellar masses in the range 0.1-100~\Msun.


\section{Sample selection and data reduction}
\label{sec:data}

\subsection{Photometric data}

The United Kingdom Infrared Telescope (UKIRT) Infrared Deep Sky Survey (UKIDSS) \citep{Lawrence2007} Ultra Deep Survey (UDS) is a near-infrared survey conducted with the UKIRT Wide Field Camera in a field centred at RA = 02$^{h}$ 17$^{m}$ 48$^{s}$ and Dec. = -05$^\circ $ 05' 57'' (J2000). At the time of the fifth data release, which was used for the study of our targets \citep{Caputi2011}, these data had reached $5\sigma$ depths of $24.0$, $23.7$ and $23.9$~ AB mag for {\em J,H,} and {\em K}, respectively.

The UKIDSS UDS field has also been covered by multiwavelength photometric data, ranging from X-rays to radio wavelengths. At optical wavelengths, the UDS field has been observed by the SuprimeCam on Subaru \citep{Furusawa2008}, providing photometry in the {\it B, V, R, i,} and {\it z} bands. In addition, complementary {\it U}-band observations (PI Almaini) have been carried out with the Megacam on the Canada-France-Hawaii Telescope. At mid-IR wavelengths, the UDS field was covered by two instruments on board the {\it Spitzer Space Telescope} (SPUDS; PI Dunlop). The Infrared Array Camera \citep[IRAC;][]{Fazio2004} covered the field with filters centred at 3.6, 4.5, 5.6, and 8 ${\rm \mu m}$, and the Multiband Imaging Photometer for {\em Spitzer} \citep[MIPS;][]{Rieke2004} with filters at 24, 70, and 160 $\rm \mu$m. 
The Subaru/XMM-Newton Deep Survey \cite[SXDS][]{Ueda2008} covered the field at X-rays, up to a depth that allows us to detect quasi stellar objects (QSOs), i.e. luminosities of $\ga 10^{44}$ erg s$^{-1}$, at redshift 3.

\subsection{Spectroscopic target selection}
Our spectroscopic program (ID 088.B-0329; PI Caputi) has targeted galaxies from a parent sample of Spitzer/IRAC 4.5 $\mu$m selected galaxies over a 0.6 deg$^{2}$ sample of the UDS Field \citep{Caputi2011}. The parent sample contained 50 321 galaxies for which photometric redshifts have been derived based on spectral energy distribution (SED) fitting on 11-band photometry, namely {\it U} through $\rm 4.5 \mu m$ bands.  The sample is  80$\%$ and 50$\%$ complete for $[4.5] < 22.4$ and $<24.0$ AB mag, respectively.

Since [4.5] is a very good proxy for stellar mass at $z>1$, one selects a population of galaxies with large stellar masses when [4.5] is used as a selection criterion. To ensure the feasibility of the spectroscopic observations we used an optical magnitude cut, i.e. $V < 25$ (AB). Although this optical magnitude cut has prevented us from targeting massive galaxies with large dust extinctions at $z\sim3$,  the resulting source list still comprises about a half of the $10^{10} < M_{\star} < 10^{11}$~\Msun galaxies at $z\sim3$, and even includes some ULIRGs, as we will discuss later. As most of the UV light is produced by massive stars, this cut introduces a bias towards star forming galaxies at $z\sim3$. We placed our two masks in regions with a surface density of massive galaxies ten times higher than average at $2.8 < \zp < 3.5$, maximizing the number of galaxies per observation. Our final target sample contains 18 galaxies with \zp $> 2.5$. In Table \ref{tab:redshifts} we give the properties of our targets. We have filled the remaining slits of the masks with galaxies that are mostly IRAC selected, some of which have a 24~$\mu m$ detection, at similar or slightly lower redshifts.

\begin{table*}
\caption{Properties of our target galaxies \label{tab:redshifts} }
\begin{center}
 \begin{tabular}{lccccccc}
\hline\hline
  {\bf Source id}$^{(1)}$ &{\bf RA (J2000)}&{\bf DEC (J2000)}&{\bf V}$^{(2)}$&{\bf [4.5]}$^{(3)}$ & {\bf Mask}$^{(4)}$& ${\bf z_{spec}}^{(5)}$&{\bf Quality}$^{(6)}$\\
\hline
26113 & 02:16:49.49&-05:14:41.01&24.73$\pm0.11$&22.26$\pm0.10$& 2&  ? & \\ 
27497 & 02:16:37.35&-05:14:01.73&23.16$\pm0.10$&22.56$\pm0.10$ & 2&  ? & \\ 
30555 & 02:16:47.57&-05:12:33.97&23.45$\pm0.10$&21.69 $\pm0.10$& 2&  2.640 & a \\
31369 & 02:16:48.60&-05:12:11.51&24.45$\pm0.11$&22.97$\pm0.10$ & 2&  0.569 & em\\
33199 & 02:16:38.56&-05:11:30.00&24.94$\pm0.12$&22.06$\pm0.10$ & 2&  ? & \\ 
34743 & 02:16:38.62&-05:10:43.03&23.73$\pm0.11$&22.02$\pm0.10$ & 2&  3.186 & a \\
36718 & 02:16:50.81&-05:09:49.88&24.14$\pm0.11$&21.35$\pm0.10$ & 2& 3.209 & b \\
37262 & 02:16:51.08&-05:09:33.06&24.04$\pm0.11$&23.37$\pm0.10$& 2& 2.934 & b \\
38729 & 02:16:39.44&-05:08:51.60&24.45$\pm0.11$&23.00$\pm0.10$ & 2&  2.611 & a \\
86032 & 02:17:19.47&-04:48:09.77&24.35$\pm0.11$&21.86$\pm0.10$ & 1&   2.696&  c \\
86588 & 02:17:29.47&-04:47:51.52&24.98$\pm0.12$&22.25$\pm0.10$ & 1&  ? & \\ 
89190 & 02:17:19.80&-04:46:22.59&24.99$\pm0.12$&22.18$\pm0.10$ & 1&  ? & \\ 
92077 & 02:17:33.44&-04:44:44.35&23.57$\pm0.10$&23.07$\pm0.10$ & 1&  $>$3.930$\tablefootmark{$\dagger$}$ & c \\
93334 & 02:17:32.06&-04:43:59.80&24.40$\pm0.11$&23.14$\pm0.10$ & 1&  ? & \\ 
94370 & 02:17:35.01&-04:43:18.91&24.40$\pm0.11$&22.84$\pm0.10$ & 1&  2.076 & em/c \\
94971 & 02:17:35.57&-04:42:59.59&24.87$\pm0.10$&21.42$\pm0.10$ & 1&  3.100 & em/a \\
95692 & 02:17:35.95&-04:42:33.64&22.57$\pm0.10$&20.13$\pm0.10$ & 1&  3.284 & a \\
96259 & 02:17:31.46&-04:42:10.74&25.00$\pm0.12$&21.85$\pm0.10$ & 1&  ? & \\ 
\hline          
25668 & 02:16:47.93&-05:14:54.43&24.35$\pm0.11$&24.18$\pm0.10$ & 2&  3.437 & em\\
97267 & 02:17:34.97&-04:41:29.23&24.46$\pm0.11$&22.76$\pm0.10$ & 1&   3.090 & c  \\
\hline
 \end{tabular}
\tablefoot{Properties of our target galaxies. Mask fillers that appeared to have $\zs > 2.5$ are included below the line. (1) ID number used; (2) {\it V}-band AB magnitude; (3) 4.5 $\rm \mu m$ AB magnitude; (4) Mask number, mask 1 and 2 have been observed for 4 and 7 hours, respectively; (5) spectroscopic redshift obtained from our FORS2 spectra; (6) redshift quality flag: $(a)$ secure redshift, $(b)$ very likely redshift but not completely secure, $(c)$ plausible redshift but not secure, $(em)$ redshift based on a single emission line.\newline
\tablefoottext{$\dagger$}{ Because we only see the Lyman $\alpha$ forest, we are unable to determine the redshift and can only give a lower limit, see also Sect. \ref{sec:specan}.} }
\end{center}
\end{table*}

\subsection{Spectroscopic data}

The spectra analysed in this work were obtained with the FORS2 spectrograph on the VLT, using the GRIS$\_$1400V+18 grism. This grism covers the wavelength range $ 4560\ \AA< \lambda < 6000\ \AA$ (1140 to 1500 $\AA$ rest frame at $z=3$) and has a resolution of $R=2100$, with a dispersion of 0.31 $\AA$ per pixel. This resolution is higher than the typical resolution in previous galaxy studies at similar redshifts, and allows us to study profiles of absorption and emission lines in more detail.  Our spectra were obtained with two FORS2 masks.  Mask 2, containing nine targets, was observed for 7 hours in the period from 22 to 25 November 2011 and on 18 January 2012, while mask 1, containing nine targets, was only observed for 4 of the awarded 8 hours (on 16 September, 11 October, and 23 October 2012). As a result of the reduced time, the signal to noise ratio (S/N) in mask 1 is reduced by a factor of $\sqrt{2}$ per resolution element, which makes it more difficult to recognise spectral features.

\subsection{Data reduction}

The multislit images were reduced using the FORS2 EsoRex pipeline (version 3.9.6). This pipeline subtracts the bias, applies the flatfield corrections, and performs the wavelength calibration using the lamp spectrum and the standard line catalogue. To improve the wavelength calibration, we have modified this line catalogue by adding five lines (\ion{Cd}{I} $\lambda\lambda$ [4678.16, 4799.92], \ion{Ne}{I} $\lambda\lambda$ [5037.75, 5330.77], \ion{He}{I} $\lambda$ [5047.70]) and removing three ( \ion{Ne}{I} $\lambda\lambda$ [5341.10, 5881.90], \ion{He}{I} $\lambda$ [5875.62]). We also set the dispersion to 0.31 \AA\ and adapted other parameters to minimise the errors in the reduction cascade. After all calibrations, cosmic rays were removed from each individual frame using the Python routine of L.A. cosmics \citep{Dokkum2001}.

Image stacking was performed using the Python package pyfits, where we stacked all individual frames together, and considered the average of every pixel after excluding the frames with the lowest and highest values. Since the pipeline was unable to properly model the sky, we subtracted the sky by hand. The FORS2 observations were performed in nodding mode, in which the observing blocks were divided into sets of three observations where the source was placed at different vertical offsets in the slit. In this way, the effect of individual pixels or rows was removed by subtracting the average of the two complementary frames from an observation, and we used this for subtraction of the sky and CCD effects. Errors for every pixel in the 1D spectrum were calculated using a bootstrap method, where we used 1000 random sets of 21 (12) observation blocks for mask 2 (1), and fitted a Gaussian to the distribution. In the cases where the fitting routine failed, we set the error to the maximum error found over the whole spectrum.

The flux calibration has an inaccuracy of $\sim 10 \%$ due to differences in the different calibration tables, i.e. different calibration tables of the same star have a flux difference of $\sim 10 \%$. These errors only influence the normalisation and slope of the continuum, but do not affect our line modelling, which depends on relative flux measurements.

To compare the SEDs, and also in the case of low S/N spectra, we rebinned the spectra to lower resolution. For this we applied a similar approach to that used for the original spectra. We first stacked the spectra pixel by pixel, rejecting the highest and lowest value, and rebinned the spectra afterwards. To calculate the errors we then applied the same bootstrapping method, but using the rebinned values.

\section{Spectral analysis}
\label{sec:results}

As previously discussed, our observations were carried out with a higher resolution grism than most previous studies, allowing us to more accurate study line profiles. We also analyse our spectra on an individual basis rather than deriving properties for composite spectra. Our high-resolution spectral study of individual galaxies at $z\sim3$ allows us to investigate the variety of physical conditions in massive galaxies at these redshifts.

\subsection{Redshift measurements}
\label{sec:zs}

\begin{figure}
 \includegraphics[width=0.48\textwidth]{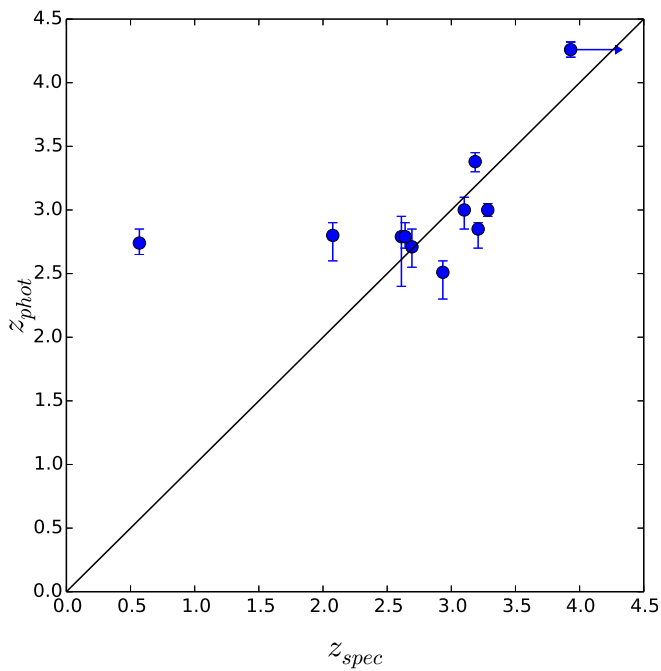}
\caption{Photometric redshifts derived in \citet{Caputi2011} versus the spectroscopic redshifts determined in this paper. Out of 11 targets with measured \zs, only one appears as a catastrophic outlier, and another one has a $\zs\sim2$, slightly lower than the photometric estimate. For 9 out of 11 targets the spectroscopic redshifts confirm the good $\zp$ estimates.\label{fig:zsvzp}}
\end{figure}

When we found multiple lines in a single spectrum, we used the centre of low-ionisation interstellar lines to determine the systemic redshift of a galaxy. For example, \ion{C}{II} $\lambda$ 1335 was used to fine-tune the redshift of galaxies 34743 and 38729.

Multiple lines of different origin were detected across our spectra. The most important and complex line in the UV is the \Lya line, which was detected in five of our spectra. Among these, two spectra show \Lya in emission, one spectrum has \Lya  in absorption, and the remaining two galaxies show P Cygni profiles. Another prominent absorption line that we detect in most spectra with a sufficiently bright continuum is the \ion{C}{II} 1335 line, which was also shown by \citet{Shapley2003} and \citet{Talia2012} to be one of the most prominent low-ionisation lines in galaxies at $z\sim 3$. We also observed that high-ionisation lines, such as \ion{N}{V}, \ion{Si}{IV}, and \ion{C}{IV}, were in general, but not always, more prominent and broader than low-ionisation lines, for example \ion{C}{II} and \ion{Si}{II}. In four spectra, we only observed a weak continuum and one or two absorption features. In these cases we decided to follow the results of \citet{Shapley2003} and \citet{Talia2012}, and assign the low-ionisation lines in the UV that were most prominent in their spectra (i.e. \ion{C}{II} and \ion{Si}{II}) to these absorption features. The non-detection of high-ionisation lines in these four spectra can be caused by the low S/N in these parts of the spectra, or can be intrinsic. For example, \citet{Talia2012} showed that \Lya emitting galaxies have relatively weaker high-ionisation lines than \Lya absorbing galaxies.

Our first goal was to measure spectroscopic redshifts for our targets. We show the results in Table \ref{tab:redshifts} and the spectra in Figs. \ref{fig:s1} to \ref{fig:sm}. For our 18 target massive galaxies with $ \zp > 2.5$, we confirmed nine to have $\zs > 2.5$, and found two galaxies with $\zs < 2.5$. For the remaining seven of our target galaxies, we were unable to determine a redshift. However, since their continuum is very faint, and they show no obvious emission lines, it is unlikely that they are low redshift sources. 

The spectroscopic confirmation that 9 out of 11 targets are at $z>2.5$ results in a success rate of $\sim$80\%. We show the comparison of photometric and spectroscopic redshift in Fig. \ref{fig:zsvzp}. In addition, we have determined spectroscopic redshifts for our mask fillers and serendipitously found that two galaxies with photometric redshifts $2.0<\zp<2.5$ have $\zs > 2.5$, resulting in a total of 11 galaxies with $\zs > 2.5$. 

\begin{figure}
\includegraphics[width=\columnwidth]{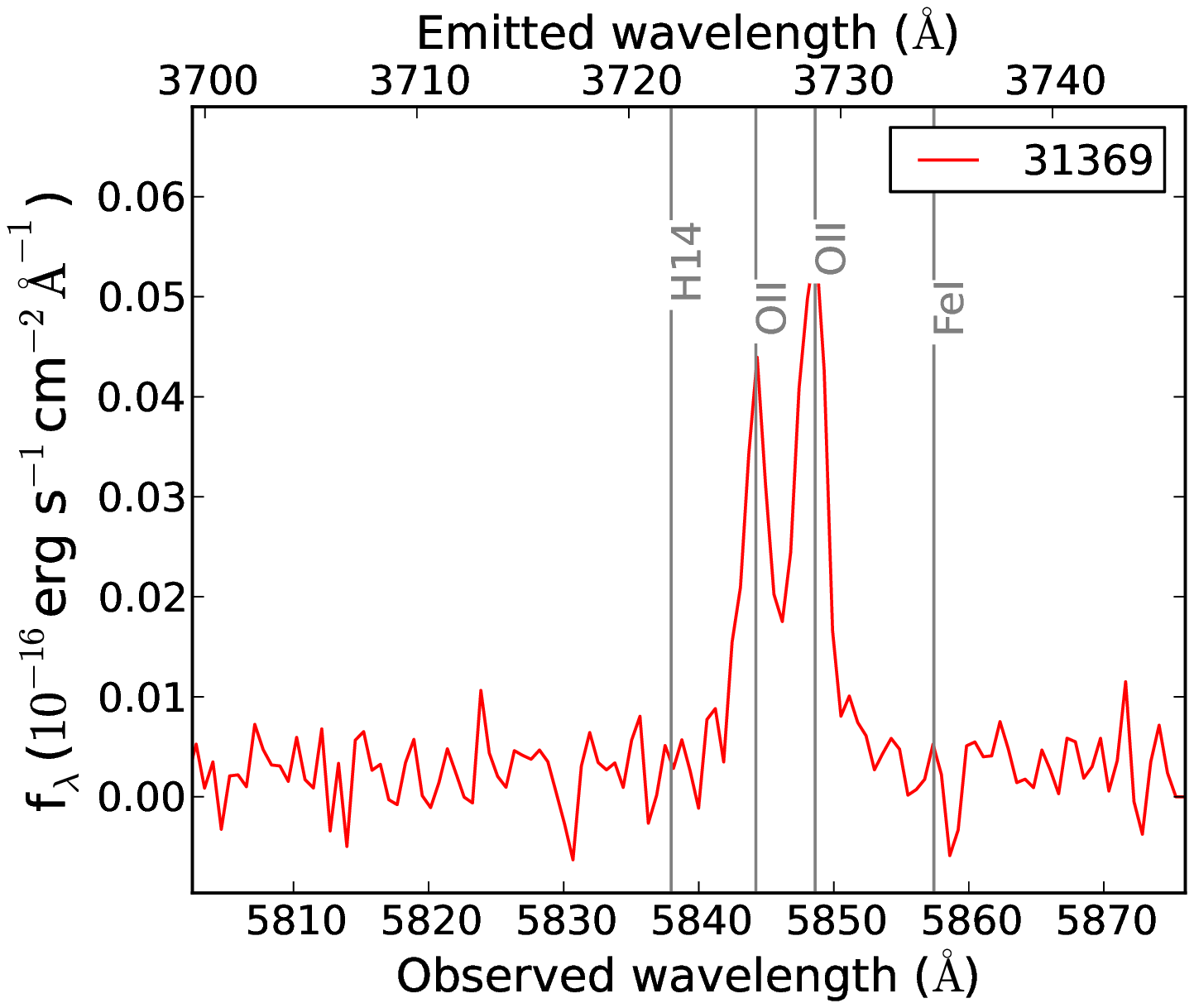}
\caption{Detailed view of the [\ion{O}{II}] emission line in the full resolution spectrum of object 31369. It is very clear at this higher resolution that there are two peaks for this emission line. \label{fig:dem}}
\end{figure}

In order to understand what causes the outliers (no. 31369 and 94370) in Fig. \ref{fig:zsvzp}, we analysed individually the photometric properties of these galaxies. In the spectrum of source 31369, we detected only a double peaked emission line (see Fig. \ref{fig:dem}). It is important to note that if we had used a similar low resolution to previous studies, it would have not have been clear that this emission line has a double peaked profile. This would have made it much harder to determine the origin of this emission line and would possibly have led to a wrong determination of the redshift. However, because the emission profile so closely resembles the profile of the [\ion{O}{II}] $\lambda\lambda$ 3726, 3729 \AA\ emission line doublet also observed in mask fillers where other emission lines provide secure redshifts, we are able to securely determine the origin of this emission line as [\ion{O}{II}]. This leads to a $\zs = 0.569$ for this galaxy, while the photometric redshift was $\zp = 2.74$. 
This spectrum shows that our higher resolution enables us to distinguish between emission lines based on their profiles. While \Lya shows a strong but broad and smooth profile, we expect to see multiple peaks for [\ion{O}{II}], \ion{C}{IV}, and \ion{C}{III}]. Therefore, identifications based on emission lines are more secure than in previous studies which used a lower resolution to determine redshifts.
The first outlier is thus caused by a strong emission line which the used SED templates do not take into account. Taking this emission line into account, and considering the correct redshift for this source, the SED models fit the broadband data well for this redshift (see also Sect. \ref{sec:models}). The high success rate for confirming high redshift galaxy candidates found in this paper and previous works indicates that
this type of outlier occurs for a small fraction of galaxies in photometric redshift samples, and they are usually difficult to prevent without the incorporation of emission lines in the templates used for SED fitting. The second outlier is a $\zs = 2.076$ source that is more than $2\sigma$ lower than its redshift estimate $\zp = 2.8^{+0.1}_{-0.2}$. The origin of this interloper is unclear, as there is no apparent contamination in the photometry.

\begin{figure*}
\includegraphics[width=\textwidth]{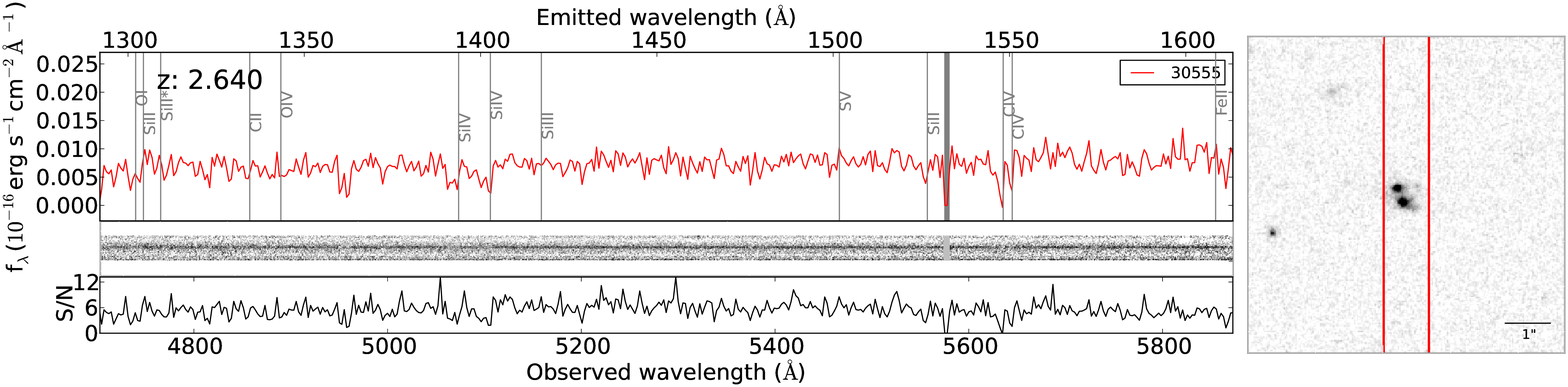}
\includegraphics[width=\textwidth]{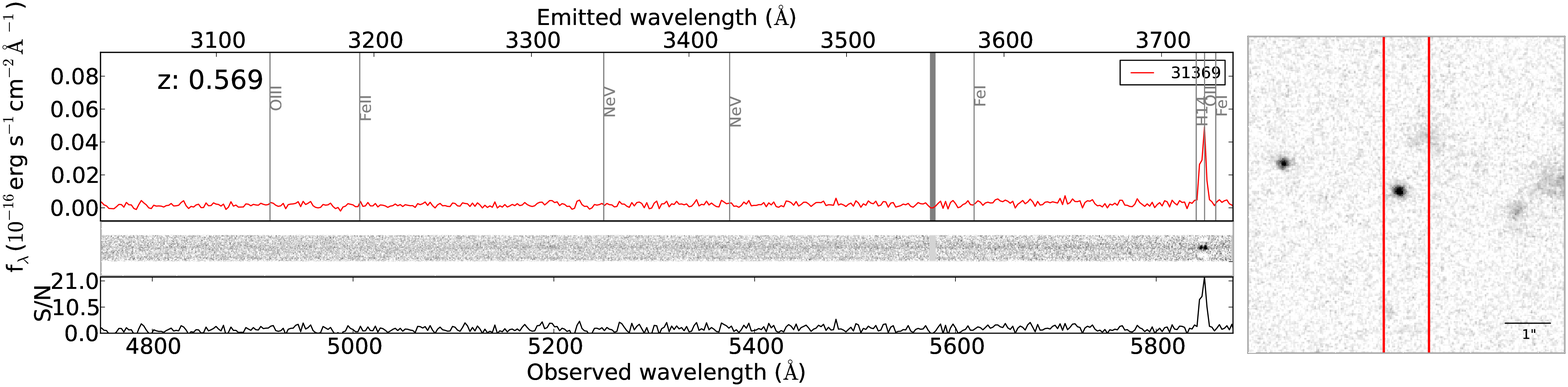}
\includegraphics[width=\textwidth]{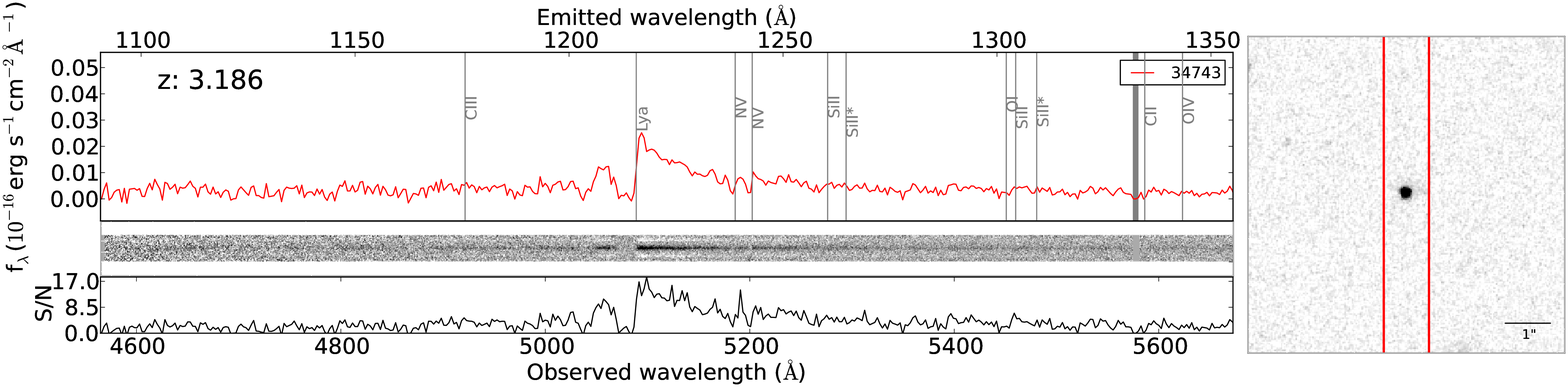}
\includegraphics[width=\textwidth]{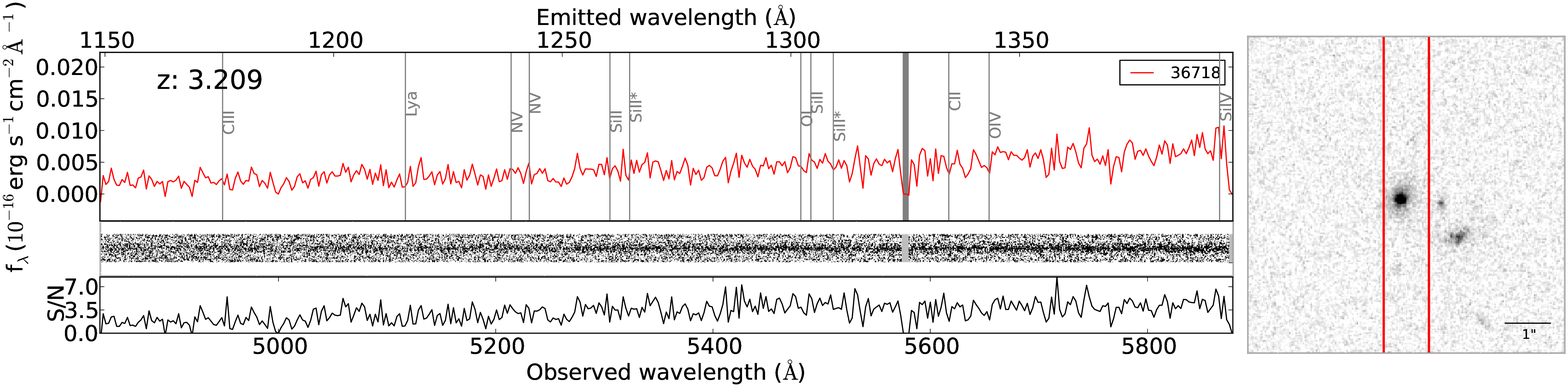}
\includegraphics[width=\textwidth]{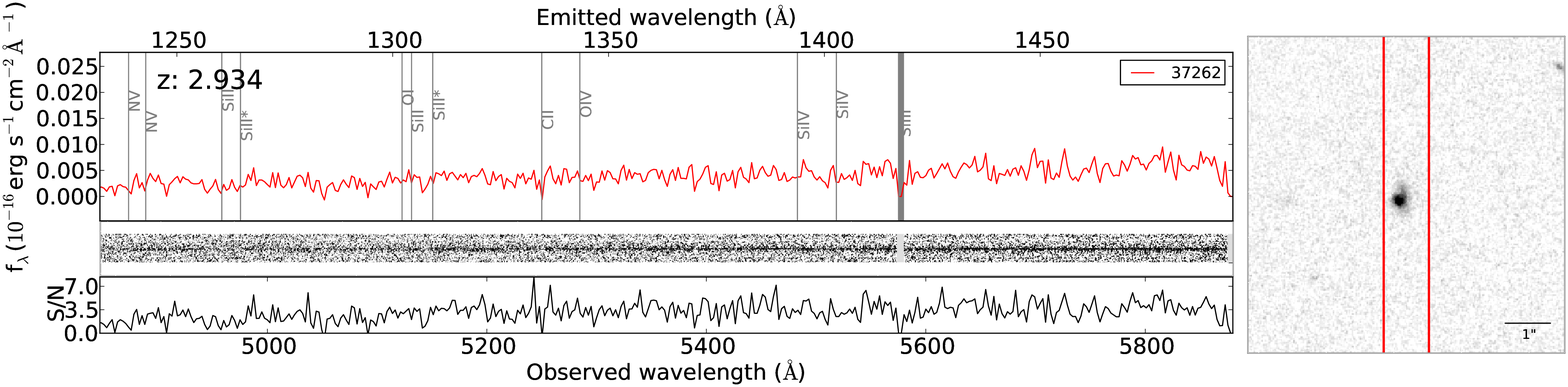}\caption{\label{fig:s1}}

\end{figure*}
\begin{figure*}
\ContinuedFloat
\includegraphics[width=\textwidth]{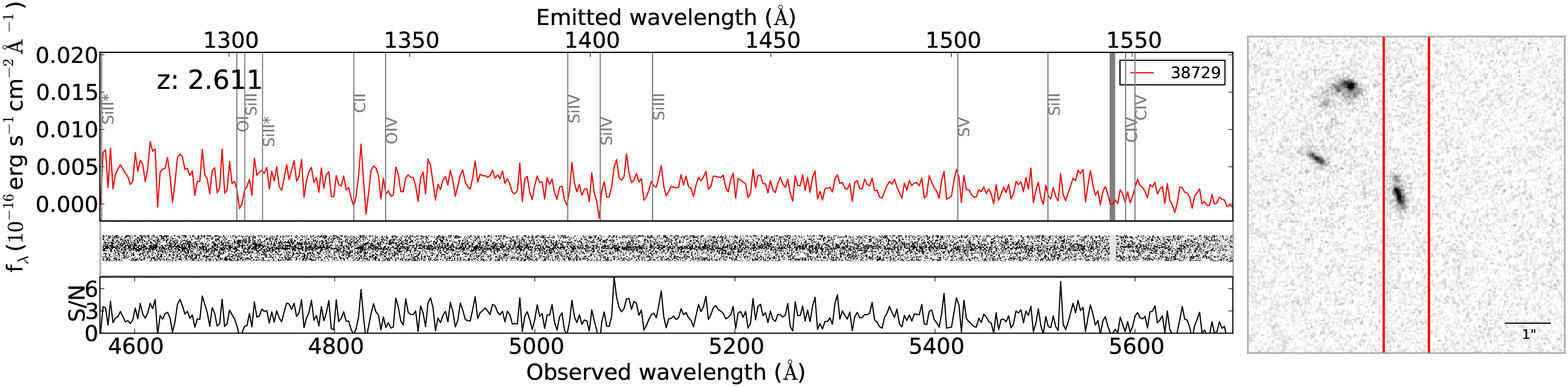}
\caption{(cont.) Rebinned spectra of targets in mask 1 with redshift determinations. The spectra are rebinned by a factor of 4 for illustrative purposes. The top spectrum is the reduced spectrum, where the top axis is rest-frame wavelength, the bottom axis the observed wavelength, and the vertical axis is the observed flux density. The determined redshift is shown in the top left corner, and the grey lines represent absorption and emission lines observed in other studies. Between 5575 \AA\ and 5580\AA\ a poorly subtracted skyline is present (marked by the grey region), where we set the spectrum to 0. The absorption features seen at this wavelength are therefore artificial. The middle plot shows the high-resolution 2D spectrum, where we subtracted the sky as explained in the text. The bottom plot shows the S/N of every wavelength element, and the stamp on the right is the $7\arcsec \times 7\arcsec$ {\em HST} ACS WFC3 stamp in the F606W filter.  \label{fig:s2}}
\end{figure*}
\begin{figure*}
\includegraphics[width=\textwidth]{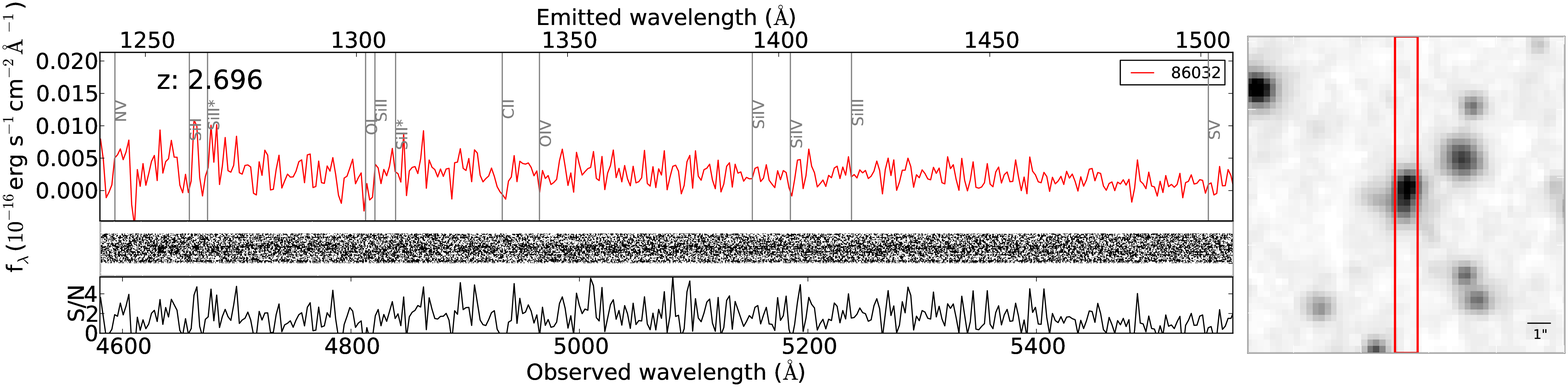}
\includegraphics[width=\textwidth]{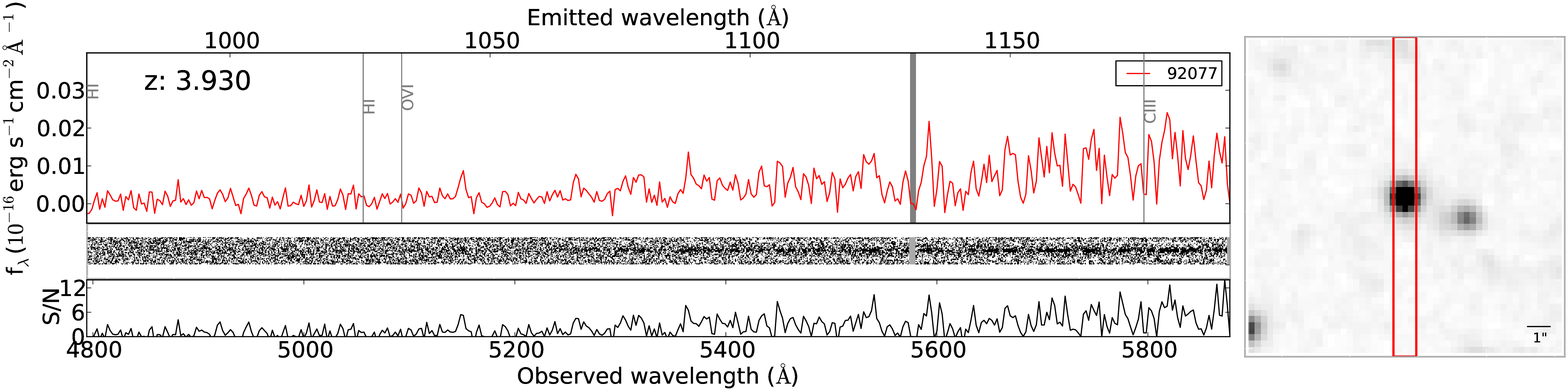}
\includegraphics[width=\textwidth]{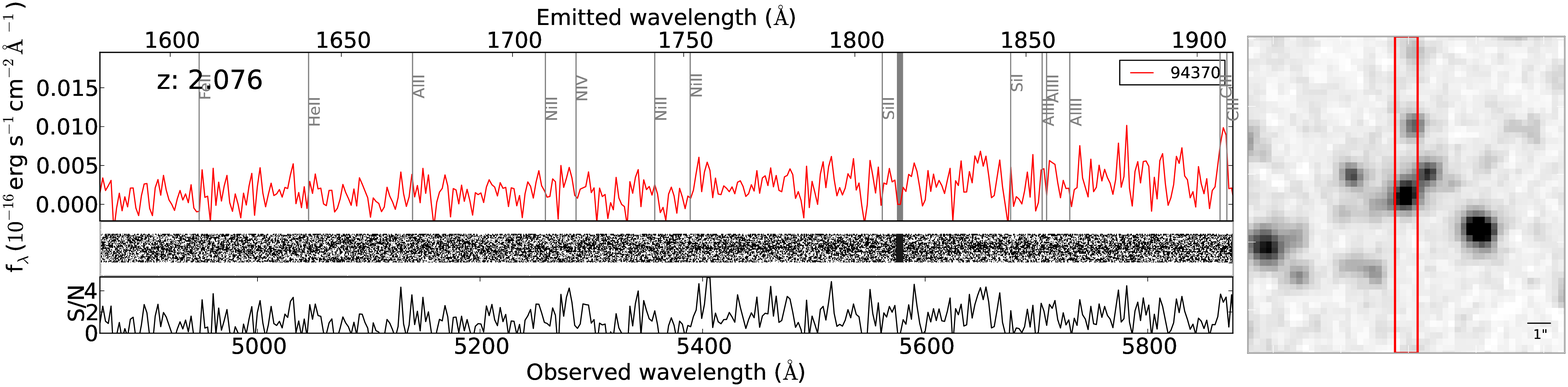}
\includegraphics[width=\textwidth]{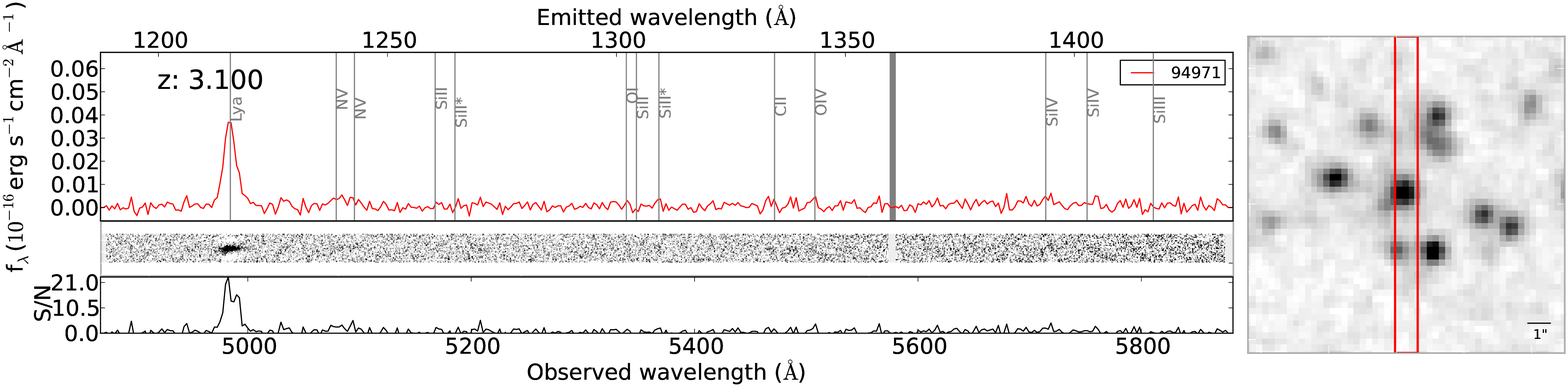}
\includegraphics[width=\textwidth]{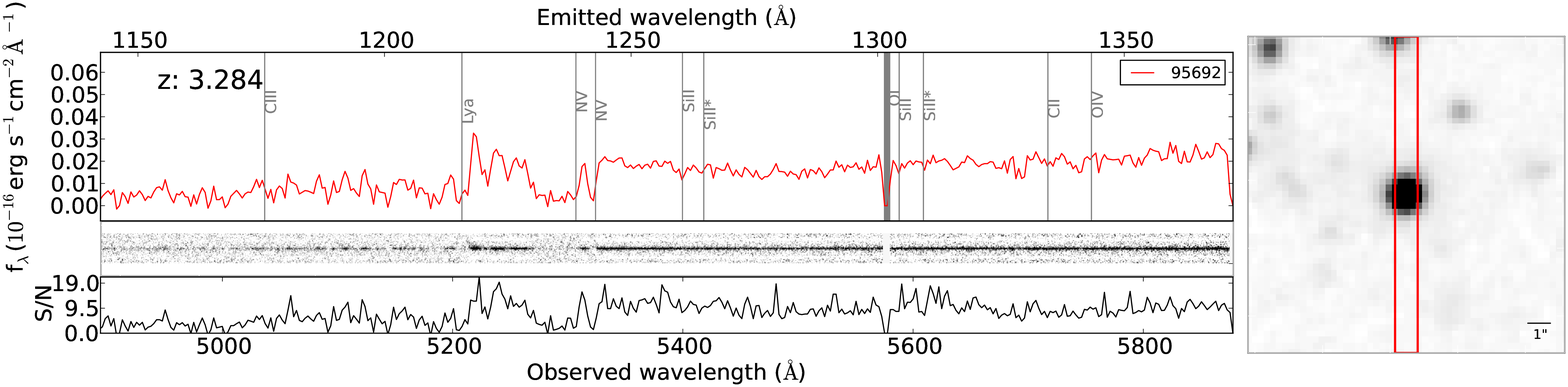}
\caption{Rebinned spectra of targets in mask 2 with redshift determinations. The axes are the same as in Fig. \ref{fig:s2}, but the stamp is from the Subaru V-band and is $14\arcsec$ on a side. \label{fig:s3}}
\end{figure*}
\begin{figure*}
\includegraphics[width=\textwidth]{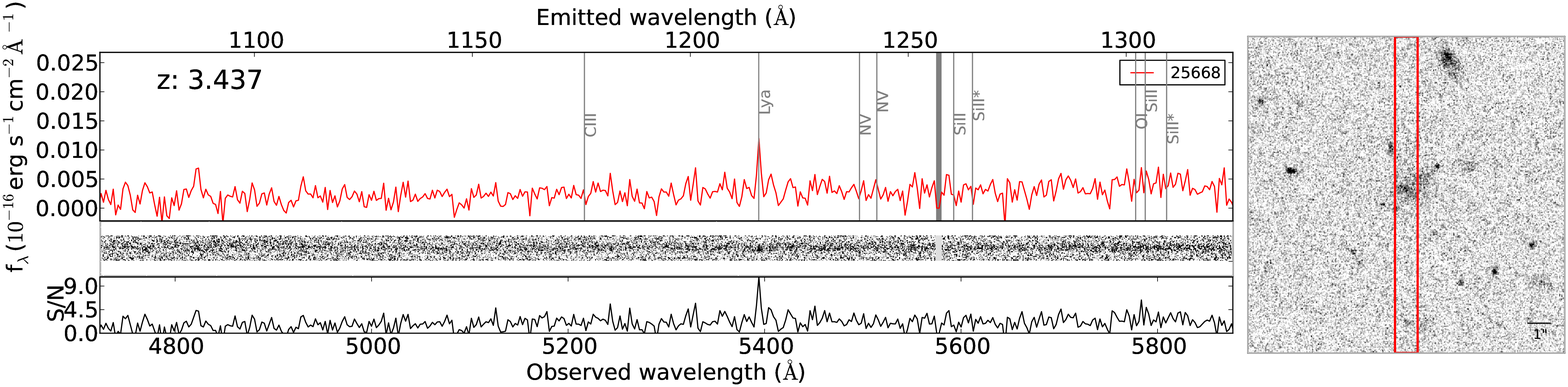}
\includegraphics[width=\textwidth]{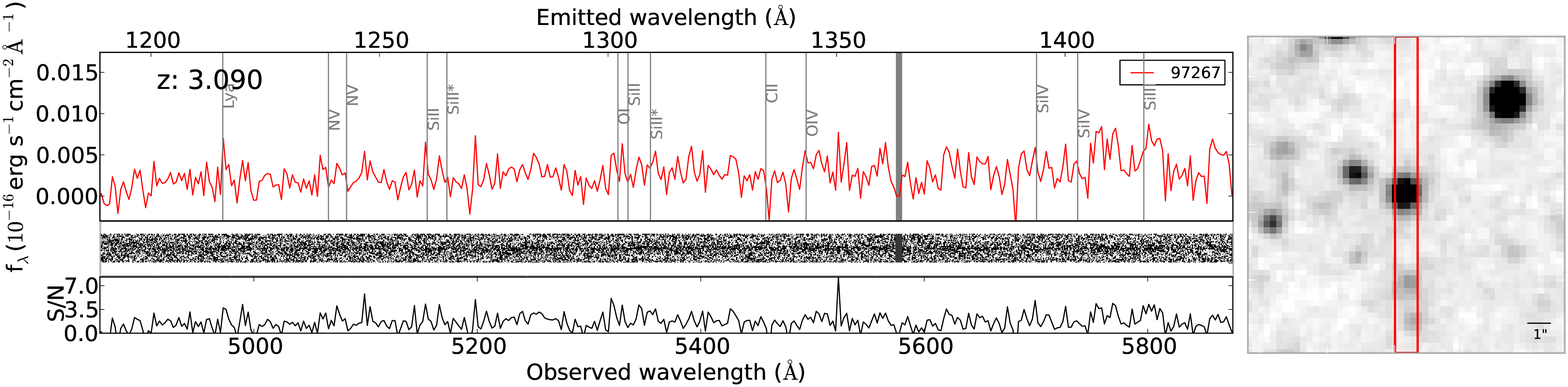}
\caption{Rebinned spectra of mask fillers from both masks with redshift determinations $\zs>2.5$. The labels and axes are the same as in Fig. \ref{fig:s2} for the top spectrum, and the same as Fig. \ref{fig:s3} for the bottom two spectra. The stamps are from {\em HST} ACS WFC3 in the F606W filter for target 25668, and from the Subaru V-band for the other two targets. The stamps are $14\arcsec$ on each side.  \label{fig:sm}}
\end{figure*}

\subsection{Analysis of individual targets with $\zs>2.5$}
\label{sec:specan}

In this section we analyse our observed spectra on an individual basis. The spectra of our targets are shown in Figs. \ref{fig:s1} and \ref{fig:s3}, while the spectra of the mask fillers are shown in Fig. \ref{fig:sm}. In all cases, we also show postage stamps of our sources: mask 2 is within the area covered by the {\em Hubble Space Telescope} Advanced
Camera for Survey ({\em HST}/ACS), as part of the CANDELS programme  \citep{Grogin2011,Koekemoer2011}.  Instead, mask 1 has no {\em HST} coverage, so we show Subaru {\em V}-band images for the sources in this mask instead.

\begin{figure*}
\includegraphics[width=\columnwidth]{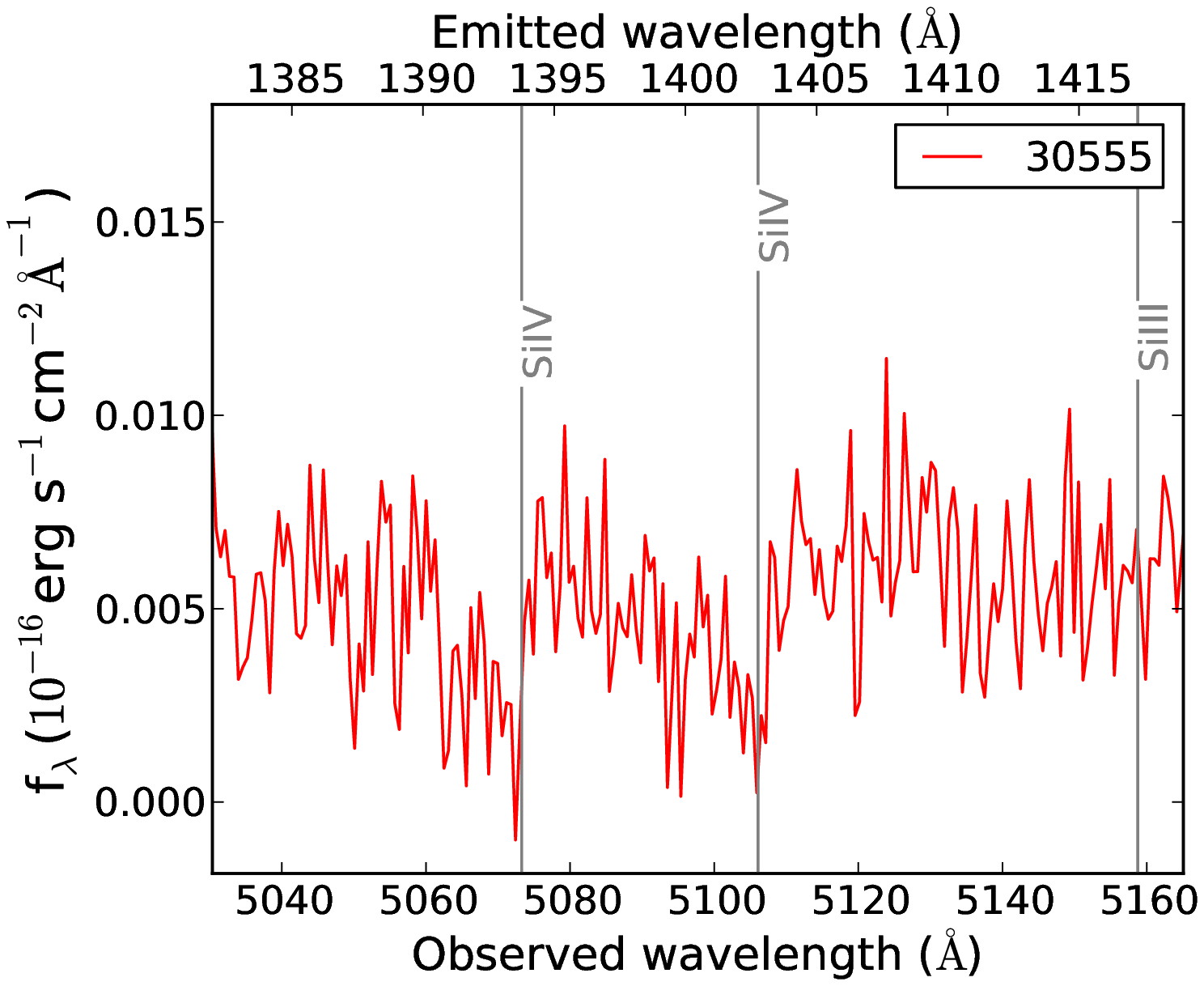}
\includegraphics[width=\columnwidth]{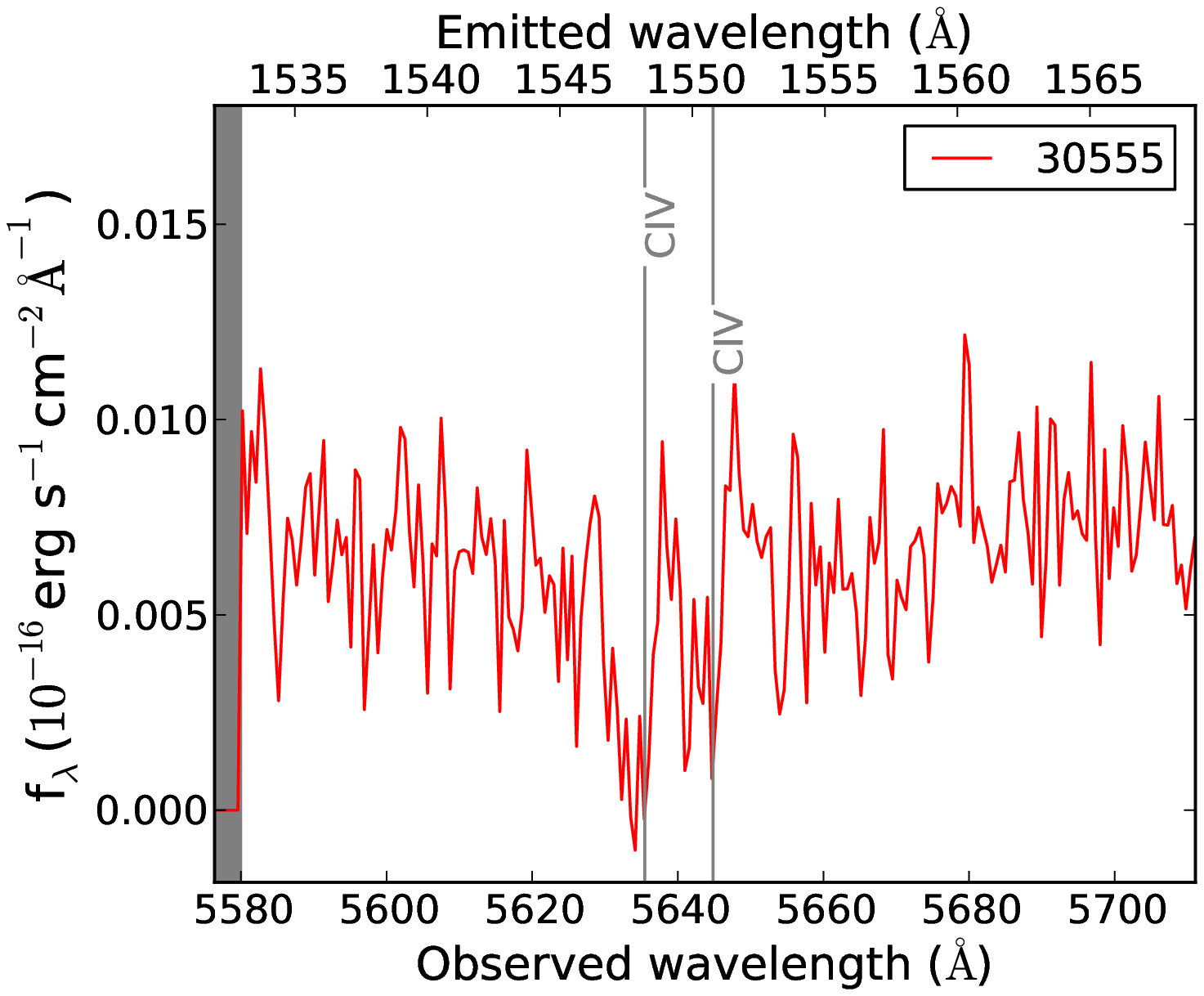}
\caption{Detailed view of full resolution spectrum of object 30555. On the left, the asymmetric profile of the \ion{Si}{IV} absorption line is clearly visible at this high resolution. On the right we show the asymmetric profile of \ion{C}{IV}; in particular \ion{C}{IV} $\lambda 1548$ \AA\ shows a clear blueshifted wing
.\label{fig:asym}}
\end{figure*}

\paragraph*{\it Object 30555:}
In the spectrum of galaxy 30555 we can clearly identify several lines, the most obvious at observed wavelengths 5071, 5104, 5555, 5635, and 5643 \AA. This combination is best fit with  \ion{Si}{IV} $\lambda\lambda 1394,1403$, \ion{Si}{II} $\lambda 1527$, and \ion{C}{IV} $\lambda\lambda 1548, 1551$ \AA, respectively, which gives the galaxy a redshift of $\zs = 2.640$. This redshift is not centred at the strongest lines, but we used the low-ionisation line \ion{Si}{II} to fine-tune the systemic redshift. We note that by using this fine-tuning, the rest-frame wavelength of the high-ionisation lines is at the wavelength with minimum flux of the absorption feature. The asymmetry of the high-ionisation absorption line profiles is studied in more detail in Sect. \ref{sec:outflows}, and from this we derived an outflow velocity of $v_{\rm out}~=~480~\pm102~\kms$ (see Fig. \ref{fig:asym} for a full resolution zoom-in on the \ion{Si}{IV}  and \ion{C}{IV} doublets). We note that we do not significantly detect \ion{C}{II} $\lambda 1335$, although there is a very weak detection visible in the spectrum. 

There is also a strong double absorption line at 4960 \AA, but there is no line combination that can also fit this feature. To see if there was some light contamination for this source, we used the available Subaru and {\em HST} images. In the Subaru bands we did not see any irregular structure of the source, but there are several other sources at small distances, whose circumgalactic media (CGM) may be responsible for these absorption lines. In the {\em HST} images (see the {\em HST} stamp in Fig. \ref{fig:s1}), however, we can distinguish two separate sources at small separation, with a slightly disturbed morphology probably caused by interaction. 

Because of this likely interaction, high SFRs are expected, consistent with the outflow velocities we have determined (see Sect. \ref{sec:outflows}). The high SFR is confirmed by a 24 ${\rm \mu m}$ detection with $S(24)=105$ ${\rm \mu Jy}$, which implies that this system is a ULIRG at $z=2.640$. With the current data it is not possible to distinguish with certainty if the two galaxies are a merger and that another galaxy is causing the absorption line at 4960 \AA, or that one of these two galaxies is a foreground galaxy. However, from the morphology and the high $24 \mu$m flux we find it more likely that this is a merging pair. If this absorption line is indeed caused by the CGM of one of the other sources at small separation or an undetected counterpart, the profile fits that of the \ion{C}{IV} doublet and the redshift of the absorber would be $\zs = 2.198$.

This merger hypothesis also affects the outflow analysis. If this is a merger pair, the light from the more distant galaxy can pass through the gaseous halo or disc of the galaxy in front of it. Owing to a velocity difference between the two galaxies, this could have a similar effect on the absorption profile as an outflow. However, the symmetric profile of \ion{Si}{II} $\lambda$ 1527 \AA\ does not agree with this scenario, although the significance of this non-detection of asymmetry in \ion{Si}{II} $\lambda$ 1527 \AA\ is not high enough to discard the pair-driven absorption profile completely. A study that includes strong and narrow emission lines can disentangle these two possibilities with certainty.

\paragraph*{\it Object 34743:}
This galaxy shows a strong \Lya line both in absorption and emission, and a clear break in the continuum due to the \Lya\ forest. The \Lya line has a complex P Cygni profile, and the complexity of the profile makes modelling this line beyond the scope of this study. In this spectrum we also see two strong absorption lines at 5180 \AA\ and 5201 \AA\ which are matched by the \ion{N}{V} $\lambda\lambda 1239, 1243 $\AA\ doublet. There are also smaller absorption features; for example we detect a weak, but significant, feature at 5584 \AA, corresponding to \ion{C}{II} $\lambda 1335$. This absorption line has an equivalent width of 1.5 \AA, more than twice that of the largest noise feature within 50 \AA. Although it is very close to the atmospheric line at 5577 \AA, there is some continuum between the absorption line and the atmospheric line. Once more we used the low-ionisation line to fine-tune the redshift to $\zs = 3.186$. As explained in Sect. \ref{sec:outflows}, we modelled the absorption profile of the \ion{N}{V} doublet and determined an outflow velocity of $v_{\rm out}~=~652~\pm64~\kms$. The {\em HST} images of this galaxy show a single compact object.

At the position of this source, \cite{Ueda2008} found a soft X-ray flux of $f(0.5-2.0 \, \rm keV)=2.7\pm0.5 \times 10^{-15}$~erg~s$^{-1}$ ~cm$^{-2}$ (assuming a photon index of 1.8), but did not find any significant detection in any other X-ray band. The total X-ray luminosity for this object is $L(0.5-2 \, {\rm keV}) = (2.5 \pm 0.5) \times 10^{44}$~erg~s$^{-1}$, i.e. it is classified as a QSO. 

\paragraph*{\it Object 36718:}
The redshift of this galaxy is based on a broad absorption line around 5120 \AA, which we identify as Ly$\alpha$. The line is shallow and not very obvious in the 1D spectrum, but the clear rise of continuum redwards of this feature strengthens this identification. Placing \Lya in the middle of this broad absorption feature results in a redshift of $\zs=3.209$, although because of the width of the line there is an uncertainty of $\Delta z = 0.01$ in this determination.

\paragraph*{\it Object 37262:}
There is only one significant absorption feature visible in the spectrum of source 37262, at $5250$ \AA. The identification of this feature with \ion{C}{II} produces a match of other expected lines with less significant absorption features, and therefore we infer that the spectroscopic redshift is $\zs = 2.934$. Because of these other possible absorption features, we classify this redshift with a flag {\em b}, rather than a flag {\em c} which would have been appropriate for a single absorption feature.

\paragraph*{\it Object 38729:}
This galaxy shows two strong absorption lines that are best fit with \ion{O}{I} and \ion{C}{II}, and also shows absorption profiles of high-ionisation lines, although at a lower S/N. This solution gives a redshift of $\zs = 2.611$, which is within 1 $\sigma$ of the photometric redshift $\zp = 2.79$.  Again, there is an asymmetry in the absorption profiles of these high-ionisation lines and we found an outflow velocity of $v_{\rm out}~=~547~\pm81~\kms$ (see Sect. \ref{sec:outflows}).

\paragraph*{\it Object 86032:}
In this low S/N spectrum we see only one absorption line. The absorption line has a slightly higher EW than nearby noise features (3.2 \AA\ and 2.3 \AA, respectively), but in the 2D spectrum this feature is slightly wider. We identify this feature with the strongest low-ionisation line in the UV, i.e. \ion{C}{II}, such that the redshift is $\zs=2.696$. As there is only one feature visible, and the S/N is very low, we classify the redshift as uncertain, i.e. flag {\em c}. This is in very good agreement with our previously determined photometric redshift of $\zp=2.71$. There are some weaker features visible, but these are not explained by any other main absorption line, and they are more likely to be noise features. There are no {\em HST} images for the mask including source, but in the Subaru images we see that there might be a fainter second source very close to the target. 

\paragraph*{\it Object 92077:}
In this spectrum there is a clear rise of the continuum flux to longer wavelengths. There are also many absorption lines visible, and there is no clear signature of either a flattening in the continuum, or a decrease in the number of absorption lines. We considered the complete Lyman series to explain either the emission lines at 5150, 5250 and 5365 \AA\ or the large variety of absorption lines, but without success. Because of this, we believe that what we observe is a \Lya forest and that we do not observe \Lya yet, but by setting the \Lya alpha line to the last absorption line we see, we find a lower limit of $\zs \geq 3.930$. This lower limit is in agreement with the photometric redshift $\zp=4.26$. 

\paragraph*{\it Object 94971:}
The spectrum of this source shows only one strong emission feature, and no continuum (this is confirmed by inspection of the 2D spectrum). The profile of this main emission line shows an asymmetry on the red tail, which suggests that this line corresponds to \Lya. 
We determined a redshift $\zs = 3.100$ for this galaxy, which is supported by other absorption features, albeit all of them very weak.

We measured a \Lya flux of 0.56 $\times 10^{-16}$ erg s$^{-1}$ and a redshift of $z=3.100$. We obtained $L_{\Lya}$ = $4.7 \times 10^{42}$ erg s$^{-1}$ without correcting for aperture or dust effects. Correcting for the extinction found in Sect. \ref{sec:models}, we found that $L_{\Lya}$ = $1.41 \times 10^{43}$ erg s$^{-1}$. 

\paragraph*{\it Object 95692:}
The spectrum of this object is the most remarkable in our sample. In this spectrum there is a break at 5215 \AA, which we associate with the \Lya forest. We see a very broad absorption line around 5300 \AA, and a very narrow absorption line at 5322 \AA. We fit the narrow absorption line with the \ion{N}{V} $\lambda$ 1243 \AA\ feature and obtain a redshift of $\zs = 3.284$. We note, however, that this redshift does not fit the weaker absorption line seen at 5694 \AA, and that by using \ion{C}{II} to fit this line $\zs = 3.267$ is found. We favour 3.284 however, because otherwise the line profiles of \Lya and \ion{N}{V} will show excess absorption at the blue side, corresponding to inflowing material. Although this might not necessarily be a problem for Ly${\rm \alpha}$, it is very unlikely that the hot and highly ionised \ion{N}{V} is present as inflowing gas, which instead is expected to be cold. 

\citet{Ueda2008} showed that there is an X-ray detection with a flux of $f(0.5-4.5 \, keV)=1.7 \times 10^{-15}$ erg s$^{-1}$ cm$^{-2}$ at a small offset from the source. \citet{Smail2008} used the AAOmega spectrograph at the Anglo-Australian Telescope to obtain a lower resolution spectrum, and estimated a redshift of 3.292, although with low confidence. This is very close to the redshift we have obtained with a higher resolution. The resulting X-ray luminosity for 95692 is $L_X\approx 1.4 \times 10^{44}$ erg s$^{-1}$, which means that this galaxy also contains a QSO.  In the \citet{Smail2008} spectrum, the broad absorption lines observed at wavelength $\sim 5300 \, \AA$ are detected, but the large width of the line is lost because of the much lower resolution. Although the resolution is too low to determine the width of absorption lines, other clear absorption features associated with high-ionisation lines are visible.

At 24 $\mu$m this galaxy has a flux density of $S_\nu(24)=(309 \pm 10$) $\mu$Jy, which makes it a ULIRG in addition to a QSO. The photometric observations in the optical and IR bands show a rising flux with wavelength (see Fig. \ref{fig:bestfits}), which is in agreement with the X-ray classification of this galaxy as an AGN. We found an outflow velocity $v_{\rm out}~=~1518~\pm146~\kms$, which is the highest velocity outflow that we measured in our sample, but consistent with other studies on ULIRGs \citep[e.g. ][]{Rupke2005b,Rupke2005c,Martin2005} or QSOs \citep[e.g.][]{Trump2006}.

It is important to realise that the broad \ion{N}{V} absorption looks very similar to the broad absorption troughs seen in broad absorption line QSOs (BALQSOs) \citep[e.g.][]{Weymann1991,Hall2002} and mini-BALQSOs \citep[e.g.][]{Churchill1999}. The high energy required to ionise \ion{N}{} four times (77 eV) is consistent with this picture, and is significantly higher than the energy required to ionise \ion{Si}{} and \ion{C}{} three times (33 and 48 eV, respectively). \citet{Misawa2007} and \citet{Ganguly2013} investigate the fraction of QSOs that have intrinsic rather than galactic absorption, and find that 30-40 per cent of the \ion{N}{V} absorbers are intrinsic absorbers. According to their classification this source might have intrinsic absorption. It remains unclear however, whether the outflows observed around the BALQSO are also present at galactic scales. Therefore, it is uncertain if we can compare this outflow to galactic scale outflows, for example the outflows discussed in this paper. As it is uncertain, we include this object in the comparison in Sect. \ref{sec:outflows}, but caution should be taken with the interpretation of this outflow velocity.

\paragraph*{\it Object 25668:}
This is a mask filler for which we determined a redshift $\zs=3.437$. In this spectrum the emission line at observed 5395\ \AA\ is very clear (see Fig. \ref{fig:sm}). Since there is only one peak, this emission line cannot be a double peaked emission line, i.e. it cannot be \ion{C}{IV}, \ion{C}{III}], or [\ion{O}{II}]. This emission line is most likely Ly$\alpha$, resulting in a redshift of $\zs=3.437$. We note that in the spectrum there is another emission feature visible at 4822 \AA, but the profile in the 2D spectrum does not resemble the profile normally seen for emision lines (i.e. it is not round, but looks more like a blob). In addition, the S/N for the feature at 4822 \AA\ is very low, which is uncharacteristic for an emission line. The S/N at 5395 \AA\ is $>9$, confirming that this is the only detected emission line.

In both the Subaru and the {\em HST} F606W and F814W images, we see that there is an extended source visible, which in combination with the \Lya emission would make this a likely \Lya blob.

\paragraph*{\it Object 97267:}
Source 97267 shows an increase in the continuum at almost 5000 \AA. In combination with the absorption feature at 5456 \AA\ we identify the redshift of this galaxy as $\zs=3.090$, where the increase in the continuum is due to the \Lya forest, and the absorption feature is due to \ion{C}{II}. Not many other features can be seen in this spectrum and therefore we set the quality of this redshift as insecure.

\paragraph*{}
Overall, we note that we found a significant difference in the equivalent width ratio of the \ion{Si}{IV} $\lambda\lambda 1393,1403$ doublet for the spectrum of galaxy 30555 compared to the nominal ratio expected in an optically thin ISM. In such an optically thin medium, the expected ratio of  equivalent widths is EW({\rm \ion{Si}{IV}$_{1393}$}/EW({\rm \ion{Si}{IV}$_{1403}$})$~\approx~2$.  We measured an EW of 1.84 and 2.27~$\rm \AA$ for \ion{Si}{IV} $\lambda 1393$ and \ion{Si}{IV} $\lambda 1403$, respectively, such that the ratio is $0.81$, while \citet{Talia2012}, for example, found a ratio of $1.42$. The lower value of our ratio suggests that the ISM, and possibly also the outflowing material, of galaxy 30555 is optically thick. We attempted to confirm this by analysing other absorption line ratios in the spectrum, but unfortunately other lines such as  \ion{O}{I} $\lambda 1302$ and \ion{Si}{II} $\lambda 1304$ are mixed with weaker unresolved lines, hampering a reliable EW ratio determination. However, a simple visual inspection  indicates that both lines seem to have comparable EW, producing a ratio closer to $\sim1$, which would support the conclusion that this galaxy has an optically thick ISM. We note that we also see a ratio differing from the nominal value of $\sim2$ for other galaxies, e.g. the \ion{N}{V} doublet in the spectrum of 34743, inidicating that the ISM in massive galaxies at $z \sim 3$ might be optically thick.

\section{Outflow analysis}
\label{sec:outlowanalysis}

\subsection{Modelling of absorption profiles with outflowing components}
\label{sec:outflows}

In the four spectra with a maximum quality flag (corresponding to galaxies 30555, 34743, 38729 and 95692)  we find high-ionisation lines whose centres are shifted with respect  to the rest-frame predictions of atomic line databases and asymmetric profiles. In these shifted high-ionisation lines we also see extended absorption wings at the blue sides of the absorption lines.   
Both this shift and the asymmetric line profiles are signatures of outflows.

We modelled the absorption profile to determine the velocity of the outflowing material with the software package {\sc VPFIT}\footnote{http://www.ast.cam.ac.uk/rfc/vpfit.html}. This code requires a continuum estimate, which we derived by fitting a spline to the spectrum, excluding regions where absorption or emission lines are visible.

One of the advantages of using {\sc VPFIT} is that it allows us to tie spectral features, which means that properties of multiple absorption lines, such as veloat city and dispersion, can be coupled to each other. 
As a consequence, all lines associated with one dynamical component, for example the ISM, are moving at the same velocity and with a tied velocity dispersion. This provides a better physical representation of the gas in the galaxy, assuming that there is only one well-mixed component at a given velocity. 
We use two dynamical components when fitting the line profiles. First we tie all low-ionisation lines and a low velocity component of the high-ionisation lines as a rest-frame feature, assuming that this is the ISM. Second, we tie all the high-velocity components of the high-ionisation lines, which represent the outflowing component.

\begin{figure}
 \includegraphics[width=\columnwidth]{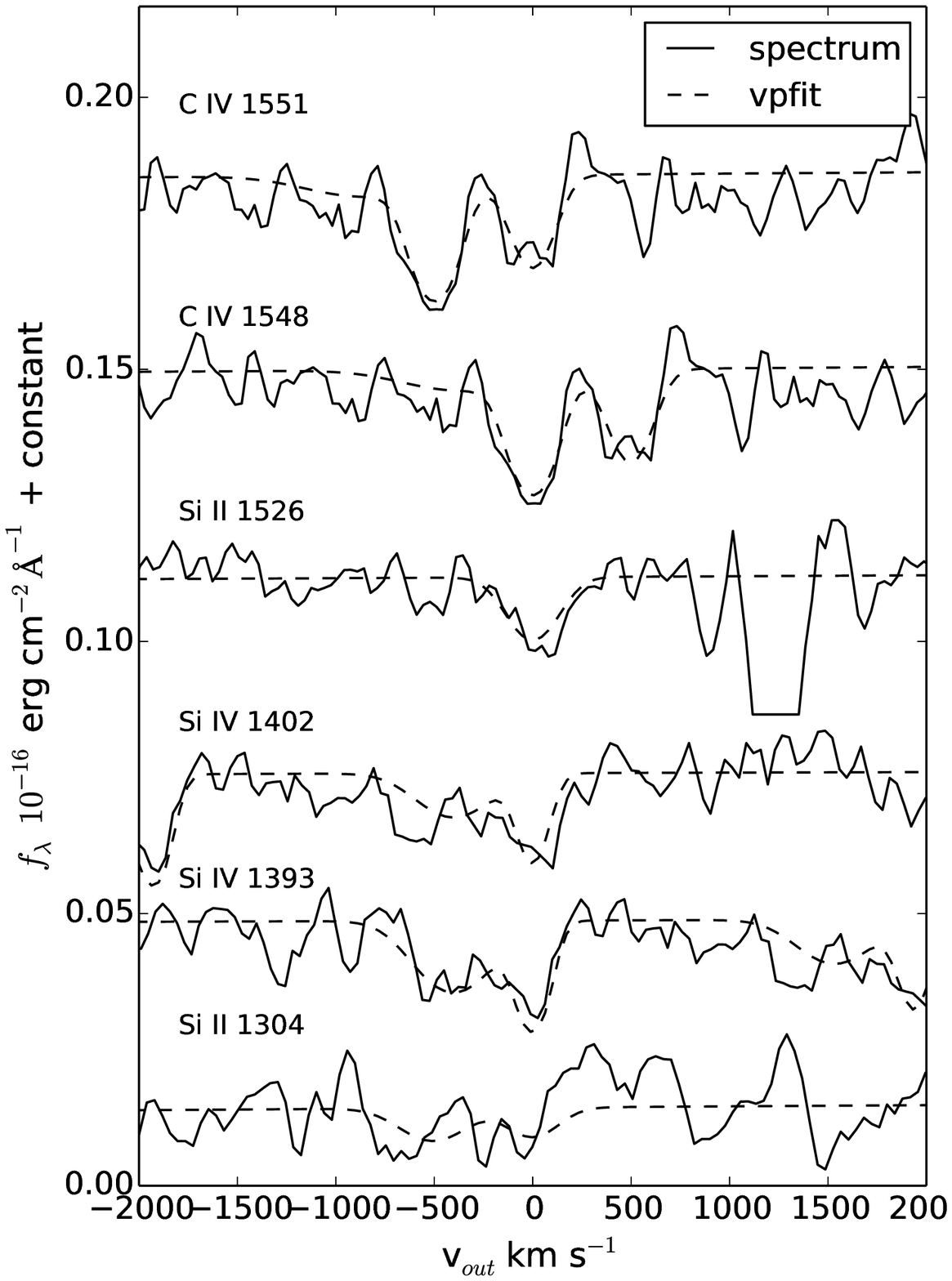}
\caption{Line profile fitting for the spectrum of galaxy 30555, obtained with the {\sc VPFIT} code. The solid line represents the spectrum, which has been smoothed with a box function of 3 pixels wide for illustrative purposes. The dotted line is the fit from {\sc VPFIT}. The horizontal axis is the velocity, where a velocity of 0 is the centre of the rest-frame feature calculated with {\sc VPFIT}, and the vertical axis is the flux plus a constant. The absorption at $\sim 500$ \kms in the \ion{Si}{II} $\lambda 1304$ \AA\ line is due to absorption of \ion{O}{I} $\lambda 1302$ \AA.}
\label{fig:out30}
\end{figure}

\begin{figure}
 \includegraphics[width=\columnwidth]{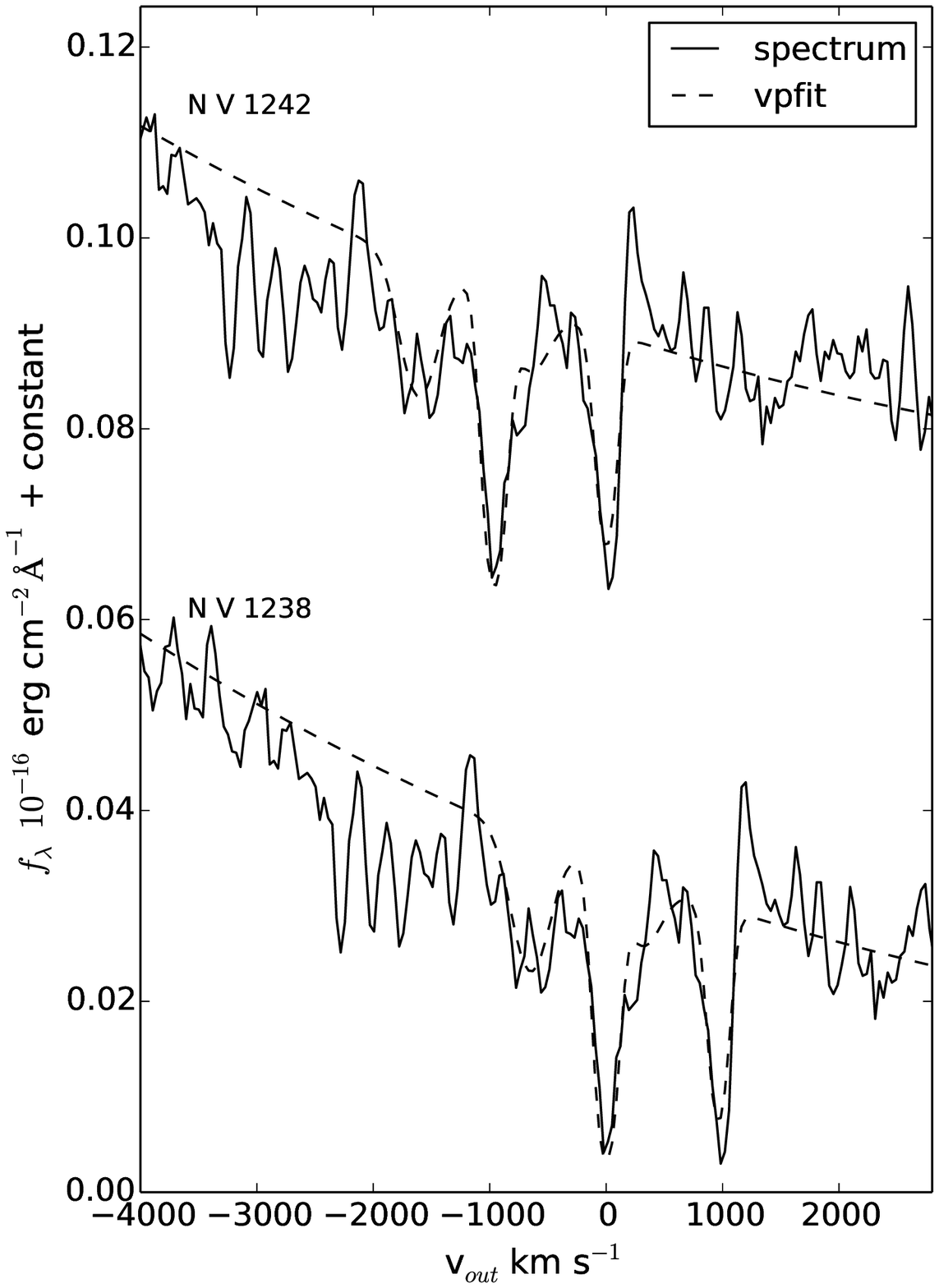}
\caption{Same as Fig. \ref{fig:out30}, but for galaxy 34743. Because of the proximity of the atmospheric line at 5577 \AA\ we did not fit the profile of \ion{C}{II}, but only used the information contained in the \ion{N}{V} line profiles. The rest-frame wavelength in this plot is therefore shifted by $\sim 300 \kms$ with respect to Fig. \ref{fig:s2}. }
\label{fig:out34}
\end{figure}

\begin{figure}
 \includegraphics[width=\columnwidth]{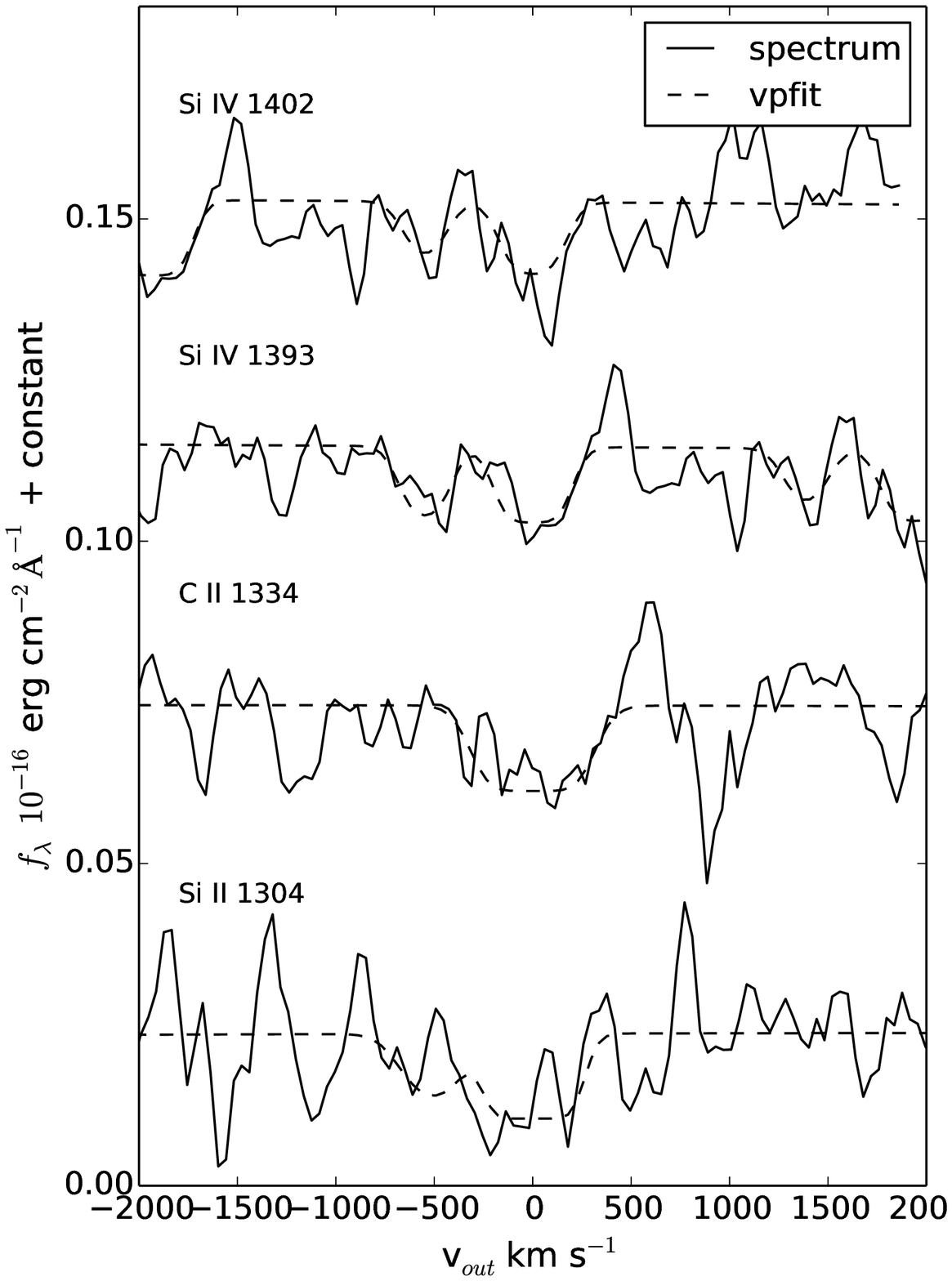}
\caption{Same as Fig. \ref{fig:out30}, but for galaxy 38729. The absorption at $\sim 500$ \kms in the \ion{Si}{II} $\lambda 1304$ \AA\ line is due to absorption of \ion{O}{I} $\lambda 1302$ \AA.}
\label{fig:out38}
\end{figure}

\begin{figure}
 \includegraphics[width=\columnwidth]{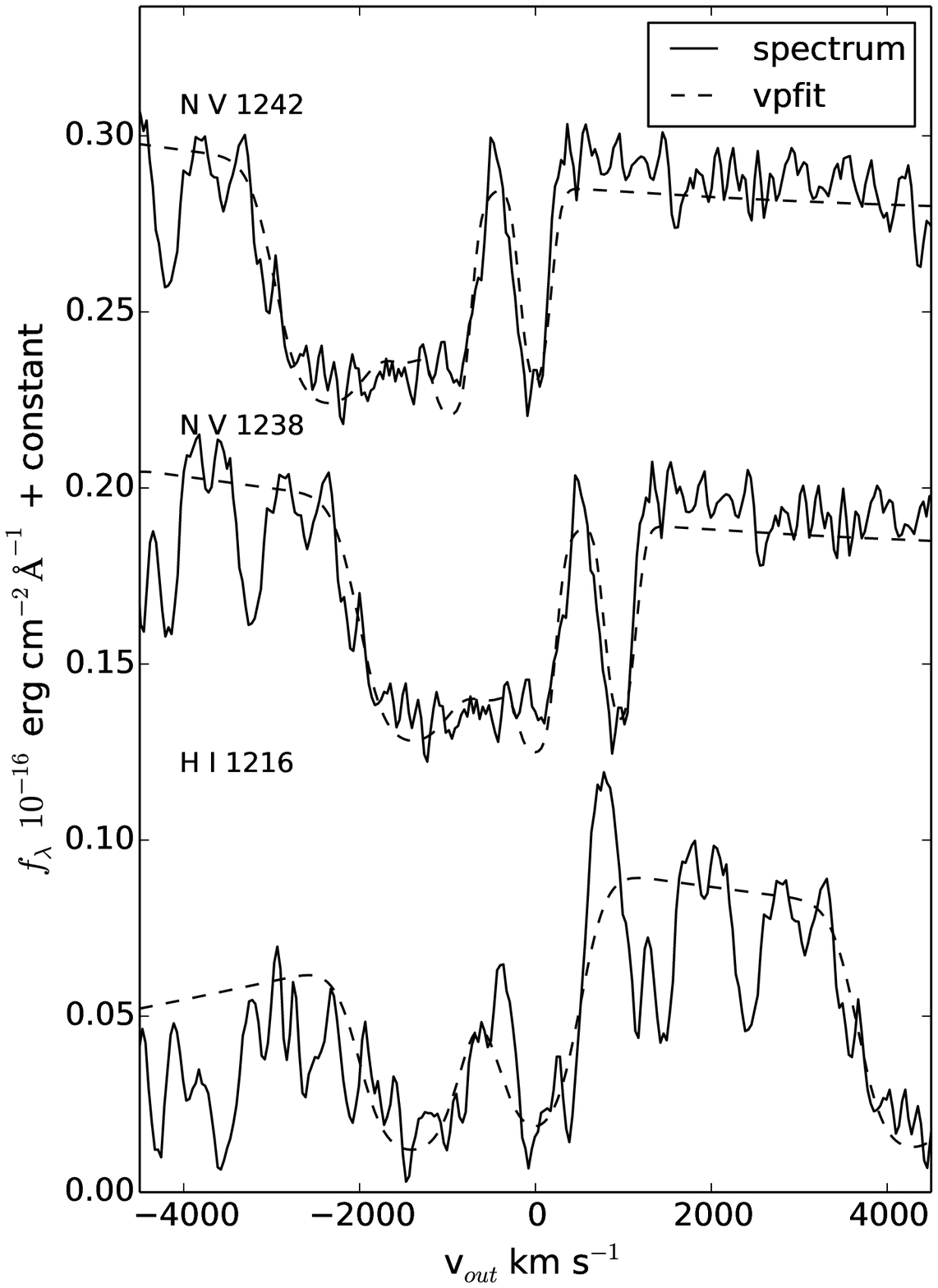}
\caption{Same as Fig. \ref{fig:out30}, but for galaxy 95692.}
\label{fig:out95}
\end{figure}

As there are only asymmetries or blueshifts in the high-ionisation lines, we use the relative difference between two components of the high-ionisation lines to measure the velocity\footnote{ Because the \ion{C}{II} absorption line in galaxy 34743 is very close to the atmospheric line at 5577 \AA, {\sc VPFIT} is unable to accurately fit this absorption profile. Rather than forcing {\sc VPFIT} to fit this line, we fitted only to the \ion{N}{V} profiles to determine the velocity.}. In two of the spectra, 30555 and 38729, we detected both the \ion{Si}{IV} and the \ion{C}{IV} features and therefore use both features to measure the outflow velocity. In both of these spectra, the asymmetries in \ion{C}{IV} are not strong enough to use this absorption line alone, but the marginal detection supports the outflow determination in \ion{Si}{IV}. The combination of the two absorption lines leads to a better redshift estimate. The results are shown in Table \ref{tab:outflows} and Figs. \ref{fig:out30} to \ref{fig:out95}. We found velocities of $\sim 600 \kms$ in three of the spectra. The fourth galaxy, 95692, displays a velocity of $\sim 1500 \kms$, more than twice the value of the other sources. A detailed analysis of this source indicates that this galaxy probably hosts an AGN, as it is a ULIRG with a 24~$\rm \mu m$ flux density of $S_\nu(24)=309 ~{\rm \mu Jy}$ and has an X-ray detection at 2-10 keV. For all of our sources we found that the component that we assumed to be at rest was at very low velocity, and at our resolution this is consistent with no velocity at all, confirming the accuracy of our redshift determinations. 

The \Lya line profile contains a wealth of information about the dynamics of the outflow. However, modelling this profile is not a simple task, as it depends on many properties, such as the column density, intrinsic line width, and velocity of the gas. As we are mostly interested in the dynamical properties of the outflow, we perform a very quick modelling of \Lya in galaxy 95692. We model the line by fitting a Lorentzian to the continuum on the blue side of Ly$\alpha$ and the continuum-subtracted profile on the red side of Ly$\alpha$. Although this is in no respect representative of the intrinsic \Lya emission, the wavelength position of the deepest absorption features will be approximately the same. As we observe an increase in the flux at 1213 \AA, we assume that there are again two dynamical components producing the absorption, a narrow absorption feature at $\sim1215$ and a broader feature $\sim1210$ \AA. We couple these \ion{H}{I} lines to the \ion{N}{V} lines as described above, and find very good agreement between atomic species (Fig. \ref{fig:out95}).

\begin{figure}
\includegraphics[width=\columnwidth]{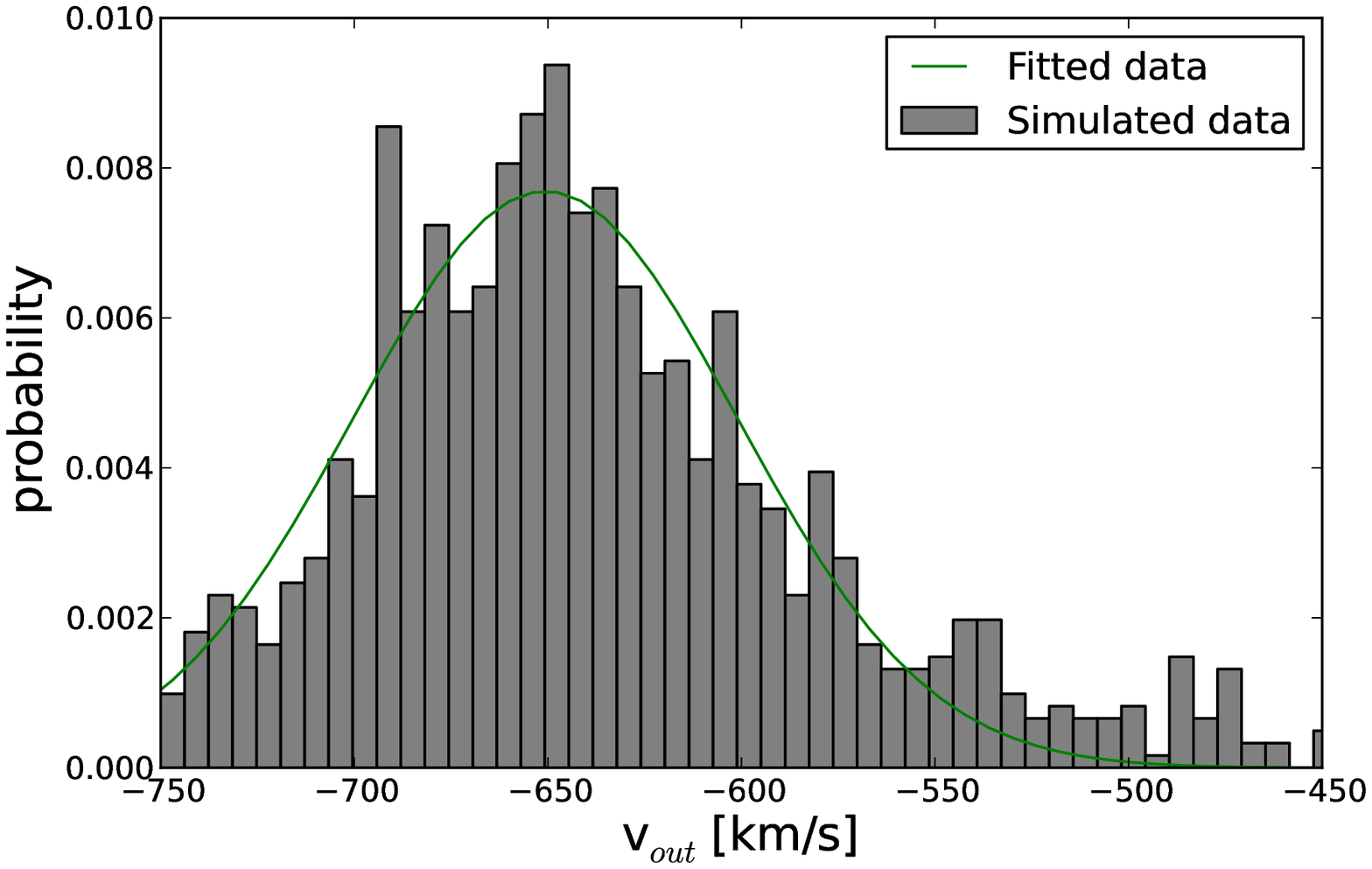}
\caption{Distribution of outflow velocities for 1000 simulated spectra based on the spectrum of galaxy 34743. The solid line represents a Gaussian fit to the resulting distribution. To obtain the final error on galaxy velocity outflow, we considered the r.m.s. of this distribution, and added an extra error component taking into account the spectral resolution.} \label{fig:histo}
\end{figure}

To determine the accuracy of this approach to measuring outflow velocities, and to see how sensitive our results are to statistical variations in the input spectrum, we created 1000 simulated spectra for each galaxy, and applied the same outflow measurement technique to them. The simulated spectra are created by assigning to each pixel a random flux density within a Gaussian distribution centred at the original flux density ($\mu$), with an r.m.s.  $\sigma$ equivalent to the measured flux density error at that wavelength. As an example, the outflow velocity distribution obtained from the mock spectra for source no. 34743 is shown in Fig. \ref{fig:histo}. We considered that the final error on the outflow velocity of each galaxy is the sum of the dispersion obtained from this distribution, and an additional error corresponding to the minimum resolution of the spectra.

\begin{table*}
\caption{Equivalent widths and outflow velocities for the four galaxies with highest S/N absorption lines. \label{tab:EWV}}
 \begin{center}
  \begin{tabular}{c|ccccccc|c}
\hline\hline
   {\bf Source ID} & \multicolumn{7}{|c|}{{\bf EW}} & ${\bf v_{\rm out}} $\\ 
 & \multicolumn{7}{|c|}{ $\left(\AA\right)$} & $(\kms)$\\ \cline{2-8}
& \ion{N}{V}$_{1238}$& \ion{N}{V}$_{1242}$& \ion{C}{II}$_{1335}$& \ion{Si}{IV}$_{1393}$& \ion{Si}{IV}$_{1403}$& \ion{C}{IV}$_{1548}$&\ion{C}{IV}$_{1551}$& \\
\hline
30555 &-&-&0.41$\pm 0.13$&1.84$\pm 0.10$&2.27$\pm 0.19$&2.39$\pm 0.33$\tablefootmark{$\dagger$}&2.39$\pm 0.33$\tablefootmark{$\dagger$}&  480$\pm 102$ \\
34743 &1.35$\pm 0.18$&0.82$\pm0.08$&1.45$\pm0.07$&-&-&-&-& 651$\pm 64$\\
38729 &-&-&2.12$\pm 0.32$&2.54$\pm 0.33$&2.50$\pm 0.45$&0.67$\pm0.28$&0.23$\pm 0.15$& 547$\pm 81$\\
95692 &11.20$\pm0.30$\tablefootmark{$\dagger$}&11.20$\pm0.30$\tablefootmark{$\dagger$}&0.66$\pm0.10$&-&-&-&-& 1518$\pm 146$\\
\hline
  \end{tabular}
 \end{center}
\tablefoot{Column 1 contains the source ID, and Cols. 2 to 8 list the measured EW for the lines seen in the spectra. In the last column the outflowing velocities measured using {\sc VPFIT} are given, for which the errors are calculated based on the analysis of 1000 mock spectra obtained from each galaxy spectrum. \\ \tablefoottext{$\dagger$}{ blended doublet due to outflows.}}
\label{tab:outflows}
\end{table*}

We did not detect any outflows in the other spectra, but this does not necessarily imply that they are not present. In these other spectra, the S/N in the continuum and the absorption profiles is not high enough to determine asymmetries with certainty.

\subsection{Comparison with previous studies}
\label{sec:comparison}

We compared our outflow velocities with those obtained by previous studies based on optical spectra. Most previously reported outflow velocities are based on line Doppler broadening, and are often upper limits on outflow velocities. In other cases, they are based on shifts between stellar and interstellar lines, which are in many cases lower limits or averages. The lack of sufficiently high resolution in outflow analysis typically leads to underestimated outflow velocities \citep[see][for a discussion]{Bordoloi2013}.

Instead, in our high-resolution study we modelled multiple dynamical components to the profile of the main absorption lines observed in our spectra, and derived outflow velocities from the difference between these components. The relatively high resolution of our spectra has allowed us  to adopt this more refined approach.

\begin{figure*}
\includegraphics[width=0.96\textwidth]{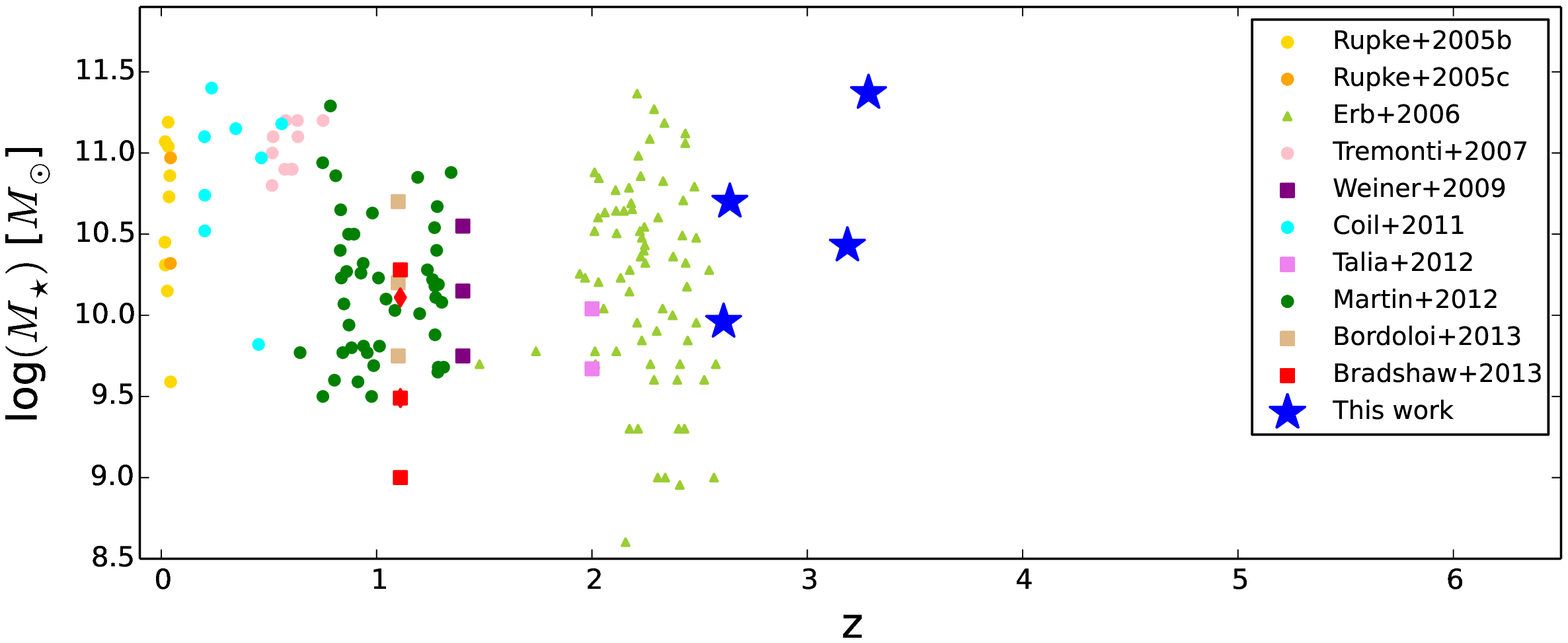}
 \includegraphics[width=0.96\textwidth]{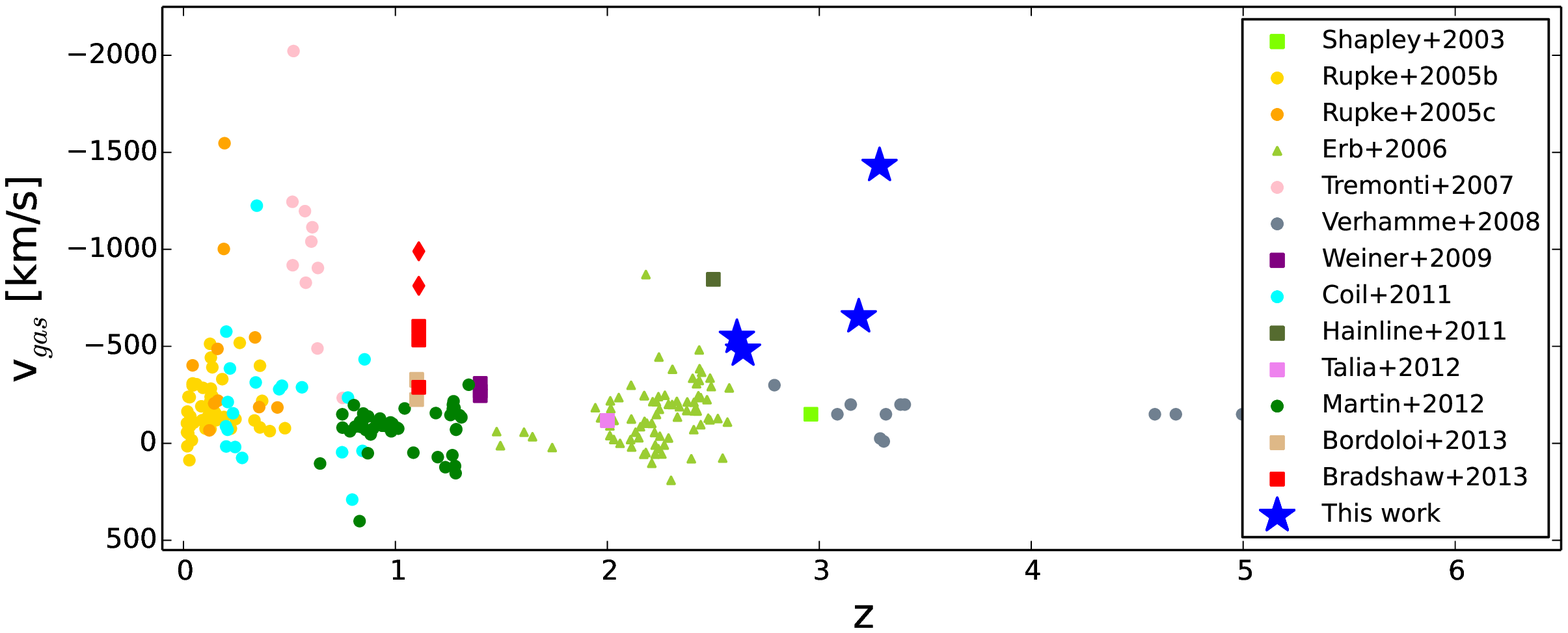}
  \includegraphics[width=0.96\textwidth]{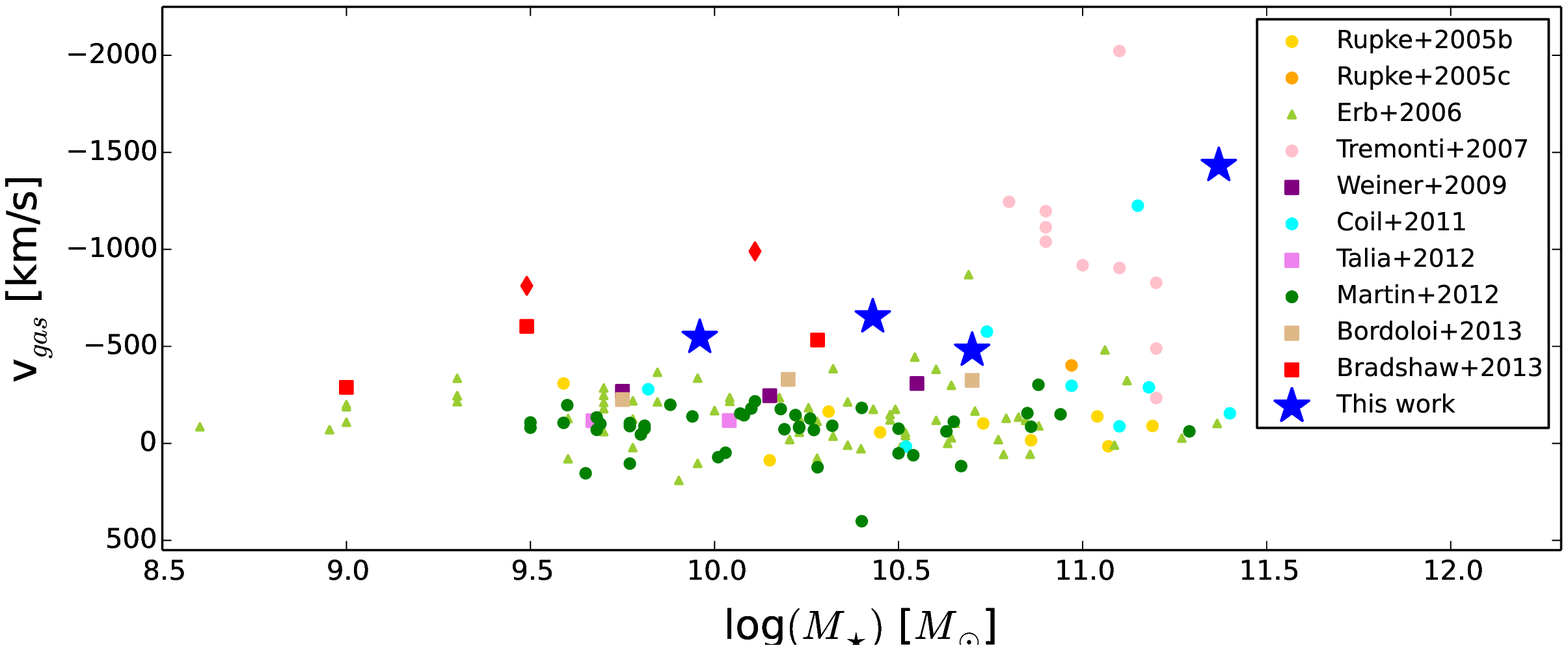}
\caption{Comparison of gas outflow velocities derived for galaxies in different studies. {\em Top:} Stellar mass versus redshift;  {\em middle:} Outflow velocities versus redshift; {\em bottom:}  Outflow velocities versus stellar mass. In the two lower plots, the $y$-axis is inverted, and a negative velocity indicates gas flowing towards us, i.e. outflows. Circles correspond to individual galaxies, squares to average measurements in composite spectra, triangles to measurements of a shift between emission and absorption lines in individual galaxies, and the star-like symbols to the results in this work. The two diamond symbols are subsamples of \citet{Bradshaw2013} containing only SF galaxies. We note that the two highest outflow velocities from \citet{Rupke2005b} ($\sim5000 \kms$ and $\sim10000 \kms$) are out of range in our plots.\label{fig:comp}}
\end{figure*}

Keeping in mind that different resolutions and methods to measure outflow velocities may lead to slightly different results, we still attempt to compare our derived values to others obtained in the literature. For this comparison, we considered studies that have adopted line profile fitting or shifts between interstellar and nebular lines \citep[i.e.][]{Shapley2003,Erb2006,Talia2012} to determine outflow velocities. Our results are shown in Fig. \ref{fig:comp}. We are clearly probing a unique region of parameter space of stellar masses and redshifts. Our stellar masses are comparable to the lower redshift surveys of \citet{Tremonti2007}, \citet{Coil2011}, and \citet{Martin2012} and the moderate redshift survey of \citet{Erb2006}. We found similar velocities to \citet{Tremonti2007} (see below for a comparison of the samples), but higher outflow velocities than \citet{Erb2006}, \citet{Coil2011} ,and \citet{Martin2012}. 

The difference in strength and occurence of outflow velocities obtained here and those in lower-redshift studies suggests that there is a significant difference in the level of feedback (due to star formation or nuclear activity) in massive galaxies at different redshifts. However, we stress that the large uncertainties in sample size and selection could influence this result and more data is needed to confirm this. If we consider that there is indeed an increase of feedback at high redshift, this is consistent with the global well-known behaviour of massive galaxy evolution,  namely that massive galaxies at low redshift are mostly quiescent, while at higher redshift many of them are actively forming stars \citep{Caputi2006b,Papovich2006,Rodighiero2010}.

\cite{Bradshaw2013} found a correlation of outflow velocity with SFR and SSFR at $z=1.1$. We calculated the SFR of all our galaxies with outflow velocity measurements, except 95692, which is strongly AGN dominated. To compute the SFR, we
considered the rest-frame UV luminosities sources, and for 30555 we also considered the total infrared luminosity ($L_{\rm IR}$), as it is 24~$\rm \mu m$ detected.

For source 30555 at $\zs=2.640$, we obtained $L_{\rm IR}$ making use of the $\nu L_\nu (8 \, \rm \mu m)$-$L_{\rm IR}$ conversion calibrated by \citet{Bavouzet2008}, and k-corrections based on the \citet{Lagache2005} models (the observed wavelength 24~$\rm \mu m$ corresponds to rest frame 6.6~$\rm \mu m$ at $\zs=2.640$).  We then followed \citet{Kennicutt1998} to derive the SFR:

\begin{equation}
{\rm SFR_{IR}} (\Msun\ {\rm yr}^{-1})= 1.73 \times 10^{-10} L_{\rm IR} (L_\odot).
\end{equation}

As this galaxy does not show any sign of AGN activity (see also Sect. \ref{sec:agn}) we did not have to correct for an AGN contribution. Using the observed 24 $\mu$m flux density of 105 $\mu$Jy, we derived an SFR of $\approx 300 \, \Msun \, \rm yr^{-1}$.

We also obtained the dust-unobscured component of the SFR traced by the rest UV light, again following \cite{Kennicutt1998}, 

\begin{equation}
{\rm SFR_{UV}} (\Msun\ {\rm yr}^{-1})= 1.4 \times 10^{-28} L_{\rm UV},
\end{equation}
where $L_{\rm UV}$ is now expressed in erg s$^{-1}$ Hz$^{-1}$, measured at 1600 \AA.  Using the uncorrected flux in the R-band, we found SFR$_{\rm UV} = 32\ \Msun$ yr$^{-1}$, and SFR$_{\rm tot} $= SFR$_{\rm UV}$ + SFR$_{\rm IR} \approx 330 \Msun$ yr$^{-1}$. Using the stellar mass obtained in Sect. \ref{sec:models}, we found log(SSFR) $\approx -8.18$.

Objects 34743 and 38729 are not detected at 24~$\mu$m, and therefore we used the attenuation-corrected rest-frame UV flux to estimate their SFR. Although 34743 has a significant power-law (PL) component in its spectral energy distribution (see Sect. \ref{sec:agn}), the rest-UV flux is dominated by the host galaxy light. We obtained SFR = 48~$\Msun$~yr$^{-1}$ and 46~$\Msun$~yr$^{-1}$, and log(SSFR) = -8.75 and -8.30, respectively. This means that the SFR of these two sources are comparable, but their SSFR differ by a factor of $\sim 2.8$. The measured outflow velocities for these two sources are very similar (overlapping within the error bars), and comparable with mean values obtained by \citet{Bradshaw2013} for galaxies with similar SFR at $z=1.1$.   Source 30555 has a much higher SSFR than 34743 and 38729, but a comparable outflow velocity. 

In summary, we do not see any clear correlation between outflow velocity and SFR or SSFR for our galaxies at $z\sim3$, as has been proposed by other authors at lower redshifts \citep[e.g.][]{Weiner2009,Bradshaw2013}. Of course, as our sample is very small we cannot conclude that these relations do not exist at such high redshifts, but if they do, scatter must be very important. The analysis of outflow velocities in larger samples of $z\sim3$ galaxies is necessary to probe if these relations actually exist at high redshifts. 

We also compared our sources to the local ULIRG sample observed by \citet{Rupke2005b,Rupke2005c}, and see that our outflow velocities are similar. All ULIRGs are associated with episodes of high star formation activity and, in the local Universe, they typically host an AGN. The high outflow velocities are a mere consequence of this powerful activity. It has been proven difficult to determine stellar masses for local ULIRGs because single aperture measurements are insufficient as they are often "messy" interacting systems. \citet{U2012} derived stellar masses for 64 local (U)LIRGs, including thirteen galaxies that are also studied in \citet{Rupke2005b} and \citet{Rupke2005c}. They find that  the stellar mass for local (U)LIRGs are similar to the masses we derived for our sample. In Fig. \ref{fig:comp} we use the masses derived by \citet{U2012} for the \cite{Rupke2005b,Rupke2005c} samples. 

High-velocity outflows were also found by \citet{Tremonti2007} in massive post-starburst galaxies at $z=0.6$, presumably also triggered by AGNs, because the star formation activity has ceased in these galaxies. The strong influence of AGNs is confirmed using a composite spectrum for AGNs at $z\approx2.5$ by \citet{Hainline2011}, in which they find a high average outflowing velocity of $\sim 850$ \kms.

In our galaxies, we expect that both intense star formation and nuclear activity are responsible for the high-velocity outflows. The two galaxies in which we measured the highest outflow velocities contain a significantPL component in their SEDs and have X-ray detections, indicating the presence of an active nucleus (see Sect. \ref{sec:agn}).
The high assembly rates of massive galaxies at high redshift are  consistent with this major activity, manifested through the high-velocity outflows.

Recently, detailed studies of outflows in local ULIRGs showed that starbursts can power outflows up to velocities of 1000 \kms, but that AGNs are necessary for higher outflow velocities \citep{Sturm2011,Rupke2013a,Veilleux2013}. In addition, \citet{Rupke2013a} argued that the ionised, neutral, and molecular gas phases are well mixed in low or moderate velocity outflows, but that the highest outflow velocities only occur in the ionised gas phase. This suggests that the two highest velocity outflows that we detect might be physically distinct from the lower velocity outflows, as the energy required to ionise \ion{N}{} four times is significantly higher than the energies required to ionise \ion{C}{} or \ion{Si}{} three times. The presence of AGNs in the two galaxies with the highest velocity outflows is consistent with the higher energies and the aforementioned studies.

\section{Properties based on multiwavelength broadband analysis of our targets}
\label{sec:models}

\subsection{Spectral energy distribution modelling}

Having obtained spectroscopic redshifts for a total of 11 galaxies at \zs$>2.5$, and two with \zs$<2.5$, we revised their broadband SED fitting results. The spectroscopic redshifts allow us to derive their properties based on broadband SED fitting, such as age, dust attenuation, and stellar mass, more securely. 

We fitted the spectral energy distribution of our targets over 13 broadbands from {\it U} through 8 ${\rm \mu m}$ at the corrected spectroscopic redshift of each source. We used our own customised SED fitting code with a library of synthetic galaxy templates generated with the publicly available packages {\sc GALAXEV} \citep{BC2003} and Starburst99 \citep{Leitherer1999}. We created a grid of these models varying the age between 0.05~Gyr and the age of the Universe at each corresponding redshift ($\sim 2.6$~Gyr at z=2.5). We considered different possible star formation histories (SFH), including a single stellar population (SSP),  several exponentially declining models with an {\it e}-folding time $\tau$, and a model corresponding to a constant star formation history. We included the effects of dust extinction by convolving each of the models with the \citet{Calzetti2000} reddening law. We considered a Salpeter initial mass function over stellar masses $0.1-100 \, \rm M_\odot$ and solar metallicity.

\begin{table}
 \caption{Parameter values used to create the grid of SED models.}
\label{tab:freepars}
\begin{center}
\begin{tabular}{cc}
\hline\hline
{\bf parameter } & {\bf parameter range}\\
\hline
Age & 5e7 -- 2.6e9 yrs \\
Metallicity & 0.02 (Z$_\odot$) \\
IMF & Salpeter \\
SFH & exponential, burst, constant\\
$\tau_{\rm exp}$ & 0.01 - 10 Gyr \\
A$_{\rm V}$ & 0 - 3.0  \\
b & 0.0 - 1.0  \\
$\alpha$ & 1.3, 2.0, 3.0 \\
\hline
\end{tabular}
\end{center}
\end{table}

As the results of the SED fitting obtained using the SB99 and BC03 templates are very similar, we only considered those obtained  with the BC03 library for the rest of our analysis (see Table \ref{tab:SED}). All but one of the selected targets has a best-fit SFH that is exponentially declining with a short {\it e}-folding time scale $\tau = 0.1$ or 0.01 Gyr. For these galaxies the ages vary between 0.05 Gyr and 0.5~Gyr, showing that young populations are preferred when combined with the exponentially declining SFHs.

For object 92077 we have been unable to determine a spectroscopic redshift and use the derived lower limit to derive the properties. Consequently, the resulting $\chi_{\rm reduced,min}^2$ value that is obtained is very high ($\chi_{\rm reduced,min}^2\approx40$), but this value decreases to $\chi_{\rm reduced,min}^2<10$ when a redshift $z>4.4$ is used. The stellar properties derived by using the lower limit $z=3.930$ are very similar to those when $z=4.5$ is used. The stellar mass decreased to $9.3 \times 10^{10}$ $\Msun$, while the best-fit age changed from 128 Myr to 72 Myr, illustrating that although the redshift is insecure, the stellar properties are reliable.

One object (30555) has a SFH characterised by much longer $\tau$ than the other galaxies, and an older SED best-fit age of $\sim 1 \, \rm Gyr$. These derived properties appear at odds with the fact that this source is classified as a ULIRG from its $24 \, \rm \mu m$ detection, i.e. it has an instantaneous dust-obscured SFR $> 200 \, M_\odot$ yr$^{-1}$. Such a high SFR could not have been continuously sustained for a 1~Gyr time period (we note that the stellar mass is only $5 \times 10^{10} \, \rm M_\odot$). A possible explanation for this apparent discrepancy might be that this galaxy had a more complicated SFH, consisting of an underlying stellar population formed continuously (which is dominant in the SED fitting), and a secondary burst with a higher SFR (producing the ULIRG phase).  

\begin{figure}
 \includegraphics[width=0.48\textwidth]{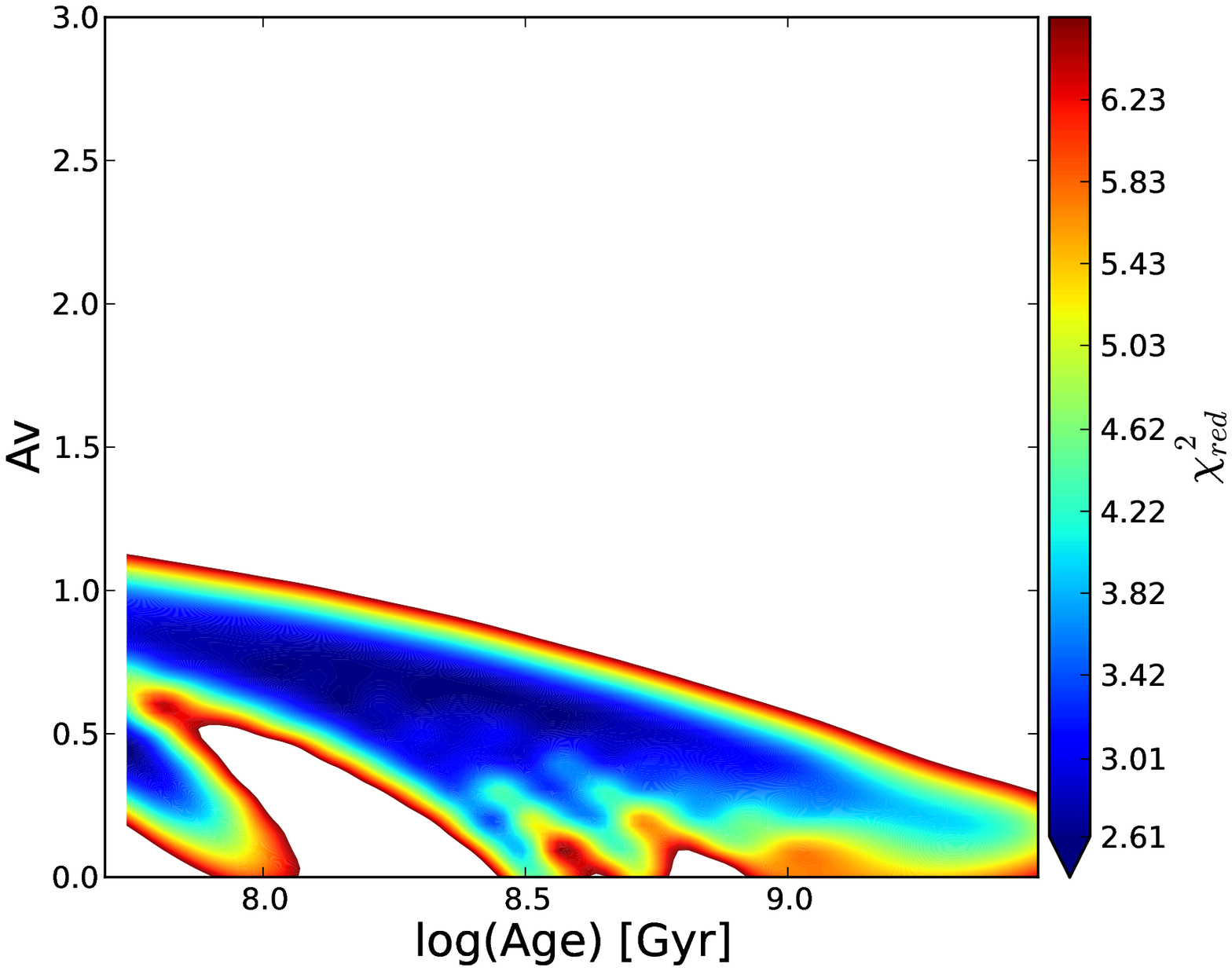}
  \caption{$\chi_{reduced}^2$ distribution of galaxy 38729. On the horizontal axis is the age in log scale, on the vertical axis is the dust extinction $A_V$. The coloured area is within $\Delta\chi_{reduced,min}^2 = 4$, or a 2$\sigma$ difference. For each grid point the best-fit model without a PL is used as the $\chi_{reduced}^2$ value. \label{fig:chiplot}}
\end{figure}

Our sample spans a range of different best-fit extinction values and other parameters. The results of the SED fitting indicate a significant degeneracy in dust extinction and age for each galaxy, as can be seen in the example in Fig. \ref{fig:chiplot}. This is a known problem, since the stellar population age and dust attenuation have similar effects on the galaxy SED. We have used our UV spectra in an effort to break or reduce this degeneracy, but the narrow wavelength coverage of our spectra, along with the relatively low resolution of the model templates,  have made this effort unsuccessful.

All except one of our galaxies have stellar masses $M_{\star}~>~9~\times~10^{9}\Msun$, including four with stellar masses $> 5 \times 10^{10}\Msun$ (after correcting for a PL contribution, when necessary; see below). These derived stellar masses indicate that almost all our galaxies are already massive at $z\sim3$. Some of them could be progenitors of the massive red galaxies that we see in the local Universe. We note that the fact that our targets only include four galaxies with stellar mass $M_{\star}> 5 \times 10^{10}\Msun$ is related to the V-band magnitude cut imposed in the spectroscopic target selection criterion, as it is known that the majority of most masssive galaxies at $z\sim3$ are very faint in the optical bands \citep{Caputi2011}.

\subsection{Analysis of the contribution of an AGN component to the SED fitting}
\label{sec:agn}

\citet{Caputi2013} proposed a generalised technique to identify the presence of AGNs among infrared-detected galaxies, and derived corrected stellar parameters, through a PL subtraction analysis. This PL component is the result of hot-dust emission produced by the AGN dusty torus, and it is manifested as excess flux at rest wavelengths $\sim 1-4 \, \rm \mu m$ with respect to the host galaxy stellar emission.

For galaxies at $z>1$,  \citet{Caputi2013} proposed the SED decomposition

\begin{equation}                                                                
S_{\nu} (\lambda) = S_{\nu}^{\rm stell.} (\lambda) +  b S_{\nu} (8 \, \rm \mu m\ ) \times \left(\frac{\lambda_{\rm obs}}{8 \, \rm \mu m}\right)^\alpha,       
\label{eq:SC}             
\end{equation}

\noindent where the term on the left-hand side is the total observed flux, and the terms on the right-hand side correspond to the stellar and PL contributions. In this equation, the PL is normalised at the observed wavelength $8 \, \rm \mu m$, which is the longest wavelength usually available to trace the maximum excess of the hot-dust emission at $z>1$; $\alpha$ is a free parameter set to values $\alpha = 1.3, 2.0, 3.0$, which are representative of the AGN spectral indices observed in the local Universe, and $b$ is the fractional contribution of the PL component to the observed total flux at 8 $\rm \mu m$. In this study we only analyse galaxies in a relatively narrow redshift range (virtually all within $2.6<\zs<3.5$), and so we considered a fixed rest-frame wavelength of $2 \, \rm \mu m$ (equivalent to observed $8 \, \rm \mu m$ at $z=3$) to normalise the PL component in Eq. \ref{eq:SC}, i.e.  

\begin{equation}
 b = 1 - \frac{S_{\nu}^{\rm stell.}(2 \, \rm \mu m\ ) }{S_{\nu}(2 \, \rm \mu m\ )}. 
\end{equation}

\noindent  This allows us a more direct comparison of the importance of the PL components in different galaxies.  The case with $b=0$ corresponds to a pure stellar template, and $b = 1$ to a pure PL at $2 \, \rm \mu m$ rest frame.  Following \citet{Caputi2013}, we determined for which of our galaxies the subtraction of a PL component from the original photometry leads to a signifiacnt improvement in the SED fitting with pure stellar templates, i.e. we imposed $\Delta\chi_{\rm reduced,min}^2 >4$. \\

\begin{figure}
\includegraphics[width=0.48\textwidth]{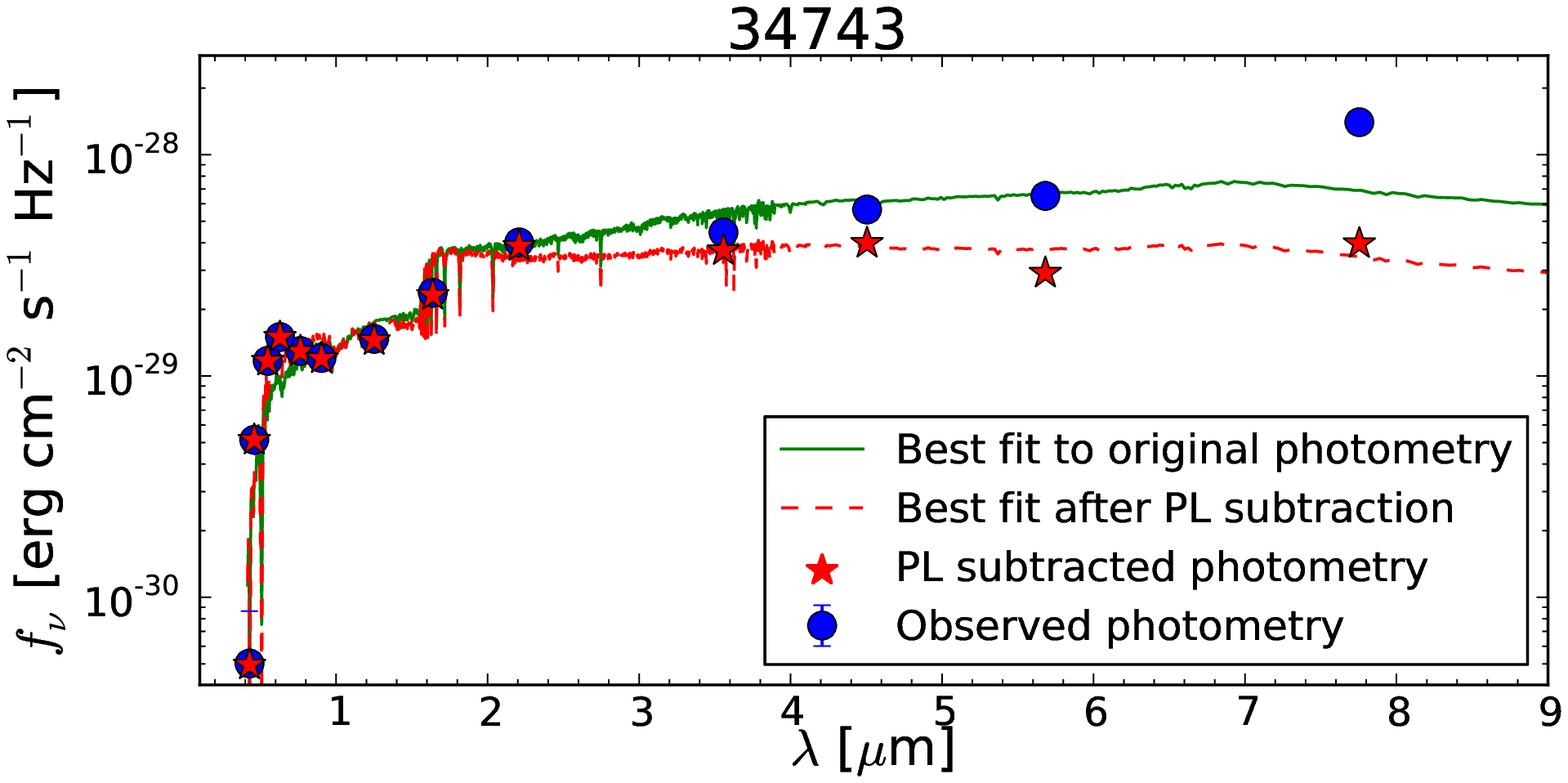}
\includegraphics[width=0.48\textwidth]{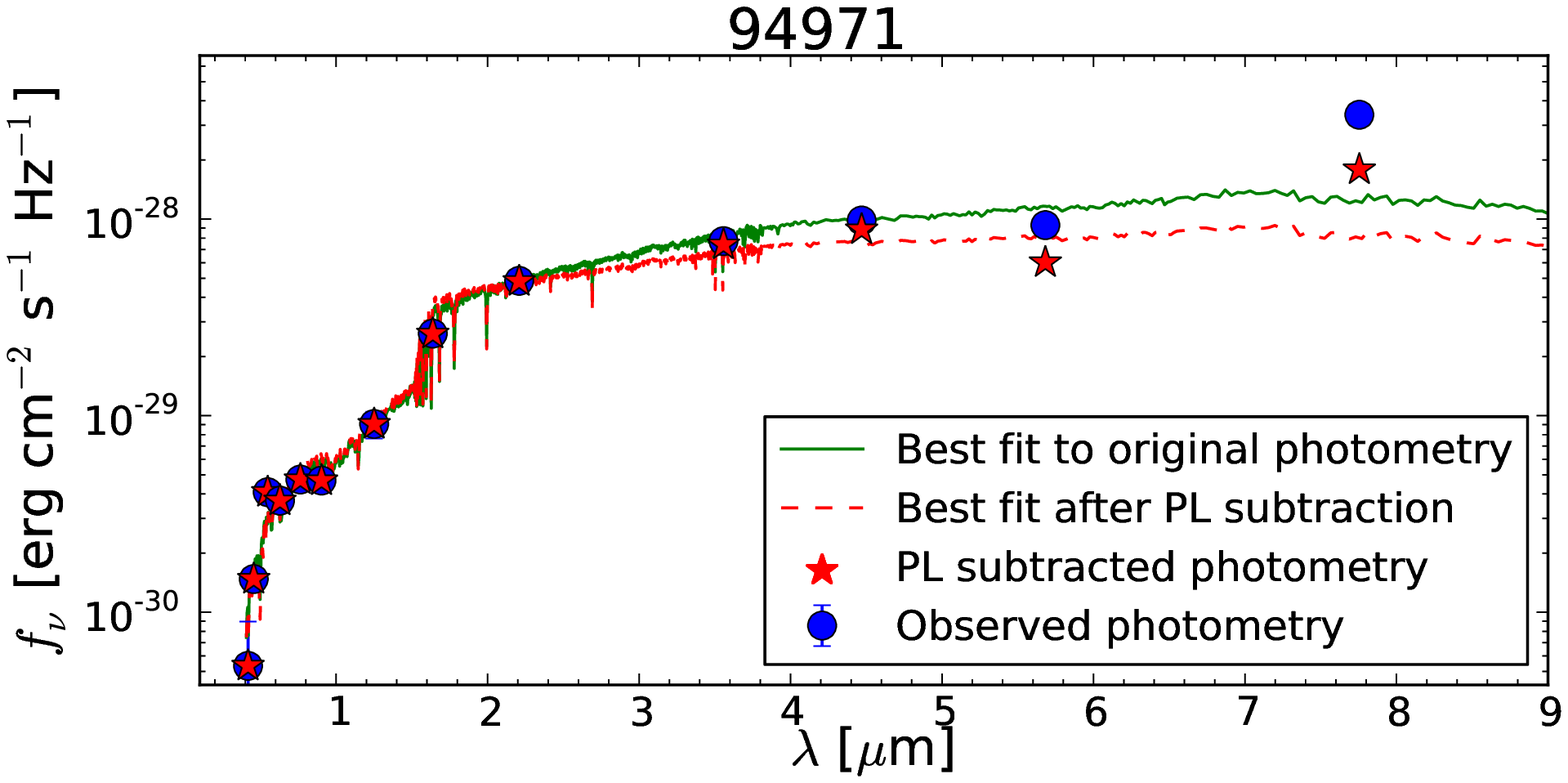}
 \includegraphics[width=0.48\textwidth]{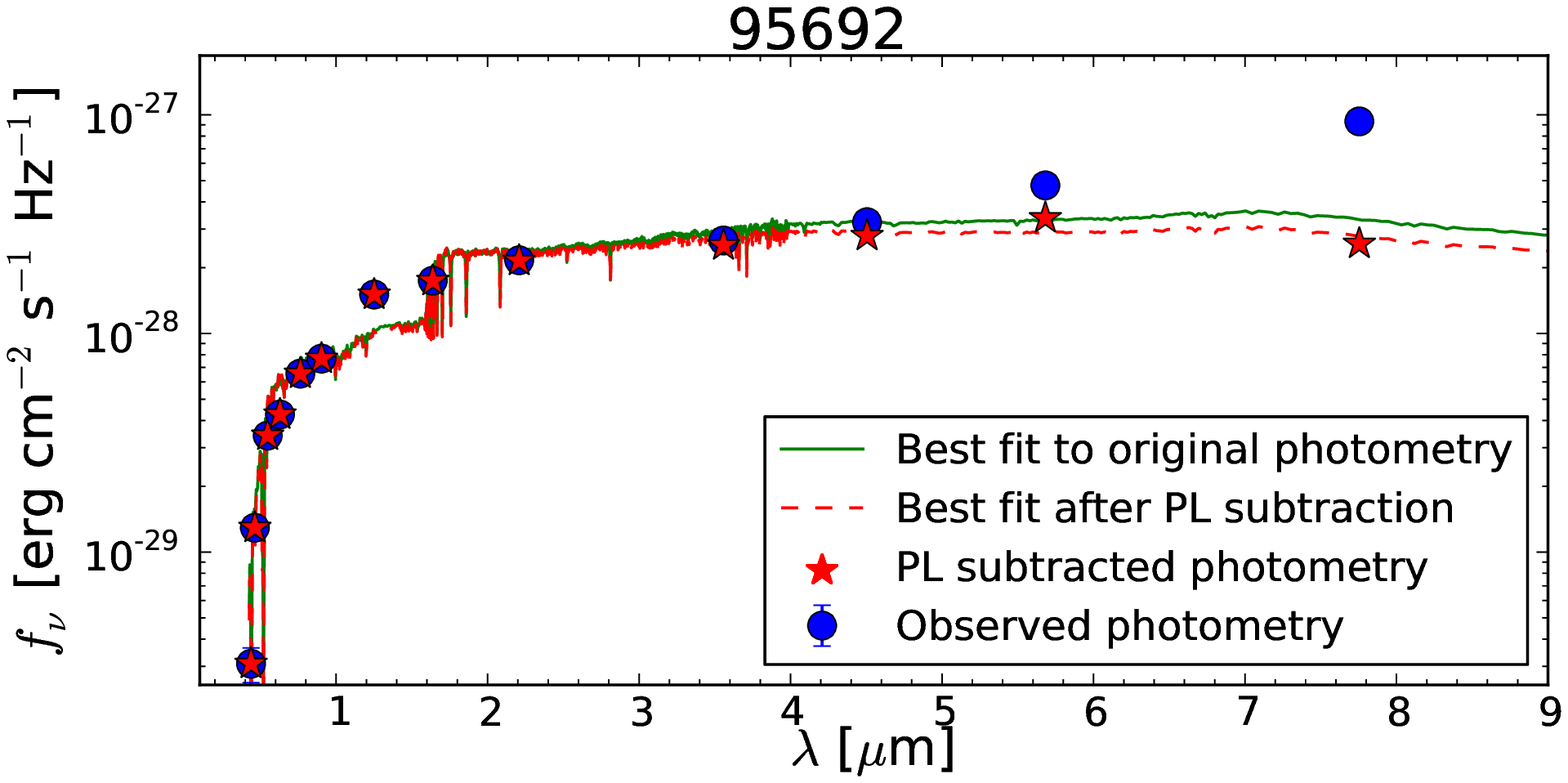}
\caption{The best-fit models for galaxies with a significant PL component, plotted with the observed broadband fluxes. The red (green) line are the best-fit models with (without) a PL component. The blue points with error bars are the observed broadband fluxes (note that most of the error margins are so small that the bars are covered by the points).\label{fig:bestfits}}
\end{figure}

Only two of our galaxies appear to have a significant PL component in their SEDs, namely galaxies 34743 and 95692, both of which are X-ray confirmed AGNs. A third galaxy, 94971, shows some excess 8 $\mu$m flux as well, but the SED fitting with and without PL subtraction gives a maximum $\Delta\chi_{reduced,min}^2 \sim 3.5$, slightly below the nominal 2$\sigma$ confidence level difference. However, we have other reasons to accept this source as an AGN. For example, this galaxy has a $24 \, \rm \mu m$  flux of 419$\pm 6$  ${\rm \mu Jy}$, indicating that a powerful dust-heating mechanism is present in this galaxy, in agreement with the flux rise already observed at 8 ${\rm \mu m}$.  In Fig. \ref{fig:bestfits}, we show the best-fit models, with and without a PL subtraction, for the three galaxies with a significant PL component in their SEDs.  As can be clearly seen, the subtraction of a PL component from the original photometry significantly improves the quality of the fitting with stellar templates. 

The best-fit stellar properties derived from the SED fitting before and after the PL subtraction are slightly different. The stellar masses decrease by a factor $\sim0.7$ for 34743 and 94971, while only a minimal stellar mass decrease is found when considering a PL for 95692. The best-fit models are more dusty for the pure stellar templates, but the ages are very similar. These effects can also be seen in Fig. \ref{fig:bestfits}, since the best-fit templates with and without a PL are very similar up to observed 4.5~$\mu$m.

\begin{table*}
\caption{Properties derived from the best-fit SED modelling of our galaxies.\label{tab:SED}}
\begin{center}
  \begin{tabular}{ccccccccc}
\hline\hline
   id\_ch2 & $z_{\rm spec}$ & SFR(t) & $\tau$ & age  &       Av &     M$_\star$ & PL & $\chi^{2}_{\rm min}$\\
   &  & &  (Gyr) & (yr)  &        &      (\Msun) &  & \\
\hline
30555&  2.640&  $e^{-t/\tau}$ &  10.000 & 1.02e+09 & 0.40 & 5.01e+10 & N & 5.45\\ 
34743&  3.186& $e^{-t/\tau}$ & 0.010 & 1.02e+08 & 0.10 & 2.72e+10 & Y & 5.99\\
36718&  3.209&  $e^{-t/\tau}$ & 0.010 & 5.50e+07 & 0.80 & 4.08e+10 & N & 6.94\\
37262&  2.934&  $e^{-t/\tau}$ & 0.010 & 7.19e+07 & 0.00 & 9.91e+09 &N& 5.76\\
38729&  2.611& $e^{-t/\tau}$ & 0.100 & 1.43e+08 & 0.60 & 9.19e+09 & N & 2.63\\
86032&  2.696& $e^{-t/\tau}$ & 0.010 & 7.19e+07 & 0.80 & 3.15e+10 & N & 4.54\\
92077&  3.930& $e^{-t/\tau}$ & 0.010& 1.28e+08 & 0.0. & 1.10e+11 & N & 39.17\tablefootmark{$\dagger$}\\
94370&  2.076& $e^{-t/\tau}$ & 0.010 & 7.19e+07 & 0.80 & 1.02e+10 & N & 4.79\\
94971&  3.100& $e^{-t/\tau}$ & 0.100 & 4.54e+08 & 0.50 & 1.06e+11 & Y\tablefootmark{$\dagger\dagger$} & 4.35\\
95692&  3.284& $e^{-t/\tau}$ & 0.010 & 1.02e+08 & 0.30 & 2.35e+11 & Y & 23.52\\
\hline
25668& 3.437& $e^{-t/\tau}$ & 0.010 & 5.50e+07 & 0.00 & 5.15e+09 & N & 5.33\\
97267& 3.090& $e^{-t/\tau}$ & 0.010 & 7.19e+07 & 0.40 & 1.77e+10 & N & 3.06\\
\hline
  \end{tabular}
\tablefoot{We used the BC03 models to construct our stellar templates, applied dust reddening according to \citet{Calzetti2000}, and used a PL component as explained in Sect. \ref{sec:models}. The parameters for the models are given in Table \ref{tab:freepars}. The Cols. are: (1) the id number of the source; (2) the spectroscopic redshift; (3) the SFH, we consider singles bursts, exponentially declining SFRs and constant SFH; (4) the free parameter for the corresponding SFH; (5) the age; (6) the dust attenuation; (7) the stellar mass; (8) the AGN classification, according to the PL component analysis; (9) the minimum $\chi^2$ value. \newline  
\tablefoottext{$\dagger$} {A lower limit is set for the redshift of this source (see text), with a high $\chi^2$ value as a result. The $\chi^2$ value decreases to $<10$ for $z>4.4$, while the properties vary slightly within a factor or 1.5.}\\
\tablefoottext{$\dagger\dagger$} {$\Delta \chi^2$ slightly lower than 4, but PL confirmation based on different properties (see text).}}
\end{center}
\end{table*}


\section{Summary and conclusions}
\label{sec:discussion}

In this work we presented the results of a VLT/FORS2 spectroscopic study of 11 massive galaxies at $z\sim3$, which have been selected from the parent IRAC catalogue by \citet{Caputi2011}. The high-redshift candidate selection was based on photometric redshifts, which we have found to be very good for identifying high-redshift sources: $\sim$80\% of the originally selected massive targets at $\zp>2.5$ were confirmed to have redshifts $\zs>2.5$.

The spectroscopic lines that we detected in our rest UV spectra are the most prominent features also found in previous studies. In our study in particular, the high-ionisation lines \ion{N}{V}, \ion{Si}{IV}, and \ion{C}{IV} are very prominent in spectra with a bright continuum. We found that, in our galaxies,  high-ionisation lines are generally more prominent than low-ionisation lines, such as \ion{C}{II} and \ion{Si}{II}. Previous studies found that low-ionisation gas is also common in outflows, but still produce the strongest absorption at rest-frame wavelengths. High-ionisation lines do not always show the strongest absorption at rest frame, but can show most absorption in the outflowing components \citep[e.g.][]{Pettini2002}. Because of this, we used the low-ionisation lines, rather than the high-ionisation lines, to determine the systemic redshifts of our sources.

Because of the relatively high resolution of the FORS2 1400V grism, we were able to resolve the profiles of emission and absorption lines better than most previous studies at similar redshifts. Using the resolved absorption profiles of high-ionisation lines, we found outflows of $>$500 \kms in four of our galaxies. 
The two highest outflow velocities are derived from \ion{N}{V} absorption line profiles. The high energy required to ionise \ion{N}{} four times and the presence of AGNs in these galaxies suggest that these outflows might be physically different from the other outflows. One of these two outflows has a very large velocity $v_{out}\sim$1500 \kms, an outflow velocity that, in the local Universe, is only found in galaxies with an active nucleus. Besides nuclear activity, the large $24 \, \rm \mu m$ flux classifies this galaxy as a ULIRG and it suggests a significant amount of star formation. To our knowledge, this is the first time that outflow velocities have been determined for individual massive galaxies at high redshift using absorption profiles in the rest-frame UV.

Our inferred large outflow velocities are in agreement with the high velocities determined for similarly massive galaxies with powerful star formation and/or nuclear activity at different redshifts. The large outflow velocities and increased levels of activity are, for example, present in local ULIRGs, which are a rare population, and constitute only a negligible fraction of the massive galaxies in the Universe today. At high redshifts, powerful star formation and nuclear activity are more widespread among massive galaxies. Our sample with outflow velocity measurements at $z\sim3$ is too small to generalise about the presence of high-velocity outflows in massive galaxies at high redshifts. Nevertheless, the fact that all four of the most secure spectra in our sample show clear high-velocity outflow signatures suggests that this phenomenon is probably quite common in massive galaxies at high redshifts. Within our sample, the high-velocity outflow incidence is $\sim 40\%$.

According to existing observational evidence, the typical level of activity among massive galaxies at $z\sim2$ should be similar to those of massive galaxies at $z\sim3$. Thus, we do not expect to observe a significant evolution in the incidence of high-velocity outflows. The differences in outflow velocities found with respect to \citet{Talia2012} can be explained by the different sample properties. As they excluded possible AGNs, they effectively removed the highest outflow velocities from their sample. In addition, the median stellar mass of their sample is significantly lower: the median stellar mass for our outfow sample is $\sim 5 \times 10^{10} \, \rm M_\odot$, while for the Talia et al. sample it is $\sim 1 \times 10^{10} \, \rm M_\odot$.

Even though inflows are expected in massive galaxies at high redshift, we have not detected evidence of them. This is not entirely surprising as their covering fraction is expected to be low and they are very difficult to observe.

Finally, we have used SED fitting techniques to re-determine stellar masses, dust attenuation and stellar ages for our target galaxies, using our  spectroscopic redshifts. We found that almost all our galaxy SEDs are best fit with stellar populations younger than 500 Myr and an exponentially declining SFH with short e-folding timescales. Eleven out of twelve galaxies have a stellar mass $M_\star > 9 \times 10^{9} \Msun$, and four have $M_\star > 5 \times 10^{10} \Msun$. The distribution of best-fit dust attenuation is broad but in all cases restricted to $Av<1$. The young ages and low to moderate dust attenuation are consequences of the selection criterion of our targets: all of them have been chosen to have $V<25$ mag to enable the spectroscopic follow up. We have also studied the presence and importance of AGNs among our targets by doing a PL component decomposition of their SED fitting. Three of our targets appear to have a significant PL component, indicative of an AGN with an underlying massive galaxy host.

Overall, this work has provided spectroscopic confirmation of the presence and nature of massive galaxies at high redshifts. In the coming years, further galaxy surveys extended to larger samples, and including redder objects, should be able to assess the general validity of our conclusions for massive galaxies at high redshifts.

\begin{acknowledgements}
 
Based on observations made with the European Southern Observatory Very Large Telescope (ESO/VLT) at Cerro Paranal, under program ID 088.B-0329. The authors thank Emma Bradshaw for providing her data points, and the anonymous referee for his/her useful report.

\end{acknowledgements}

\bibliographystyle{aa}
\bibliography{Outflows_arxiv_v4}

\begin{thebibliography}{74}
\expandafter\ifx\csname natexlab\endcsname\relax\def\natexlab#1{#1}\fi

\bibitem[{{Bavouzet} {et~al.}(2008){Bavouzet}, {Dole}, {Le Floc'h}, {Caputi},
  {Lagache}, \& {Kochanek}}]{Bavouzet2008}
{Bavouzet}, N., {Dole}, H., {Le Floc'h}, E., {et~al.} 2008, \aap, 479, 83

\bibitem[{{Bordoloi} {et~al.}(2013){Bordoloi}, {Lilly}, {Hardmeier}, \&
  et~al.}]{Bordoloi2013}
{Bordoloi}, R., {Lilly}, S.~J., {Hardmeier}, E., \& et~al. 2013, ArXiv e-prints

\bibitem[{{Bower} {et~al.}(2006){Bower}, {Benson}, {Malbon}, {Helly}, {Frenk},
  {Baugh}, {Cole}, \& {Lacey}}]{Bower2006}
{Bower}, R.~G., {Benson}, A.~J., {Malbon}, R., {et~al.} 2006, \mnras, 370, 645

\bibitem[{{Bradshaw} {et~al.}(2013){Bradshaw}, {Almaini}, {Hartley}, {Smith},
  {Conselice}, {Dunlop}, {Simpson}, {Chuter}, {Cirasuolo}, {Foucaud}, {McLure},
  {Mortlock}, \& {Pearce}}]{Bradshaw2013}
{Bradshaw}, E.~J., {Almaini}, O., {Hartley}, W.~G., {et~al.} 2013, \mnras, 433,
  194

\bibitem[{{Bruzual} \& {Charlot}(2003)}]{BC2003}
{Bruzual}, G. \& {Charlot}, S. 2003, \mnras, 344, 1000

\bibitem[{{Calzetti} {et~al.}(2000){Calzetti}, {Armus}, {Bohlin}, {Kinney},
  {Koornneef}, \& {Storchi-Bergmann}}]{Calzetti2000}
{Calzetti}, D., {Armus}, L., {Bohlin}, R.~C., {et~al.} 2000, \apj, 533, 682

\bibitem[{{Caputi}(2013)}]{Caputi2013}
{Caputi}, K.~I. 2013, \apj, 768, 103

\bibitem[{{Caputi} {et~al.}(2011){Caputi}, {Cirasuolo}, {Dunlop}, {McLure},
  {Farrah}, \& {Almaini}}]{Caputi2011}
{Caputi}, K.~I., {Cirasuolo}, M., {Dunlop}, J.~S., {et~al.} 2011, \mnras, 413,
  162

\bibitem[{{Caputi} {et~al.}(2006{\natexlab{a}}){Caputi}, {Dole}, {Lagache},
  {McLure}, {Dunlop}, {Puget}, {Le Floc'h}, \&
  {P{\'e}rez-Gonz{\'a}lez}}]{Caputi2006b}
{Caputi}, K.~I., {Dole}, H., {Lagache}, G., {et~al.} 2006{\natexlab{a}}, \aap,
  454, 143

\bibitem[{{Caputi} {et~al.}(2006{\natexlab{b}}){Caputi}, {McLure}, {Dunlop},
  {Cirasuolo}, \& {Schael}}]{Caputi2006a}
{Caputi}, K.~I., {McLure}, R.~J., {Dunlop}, J.~S., {Cirasuolo}, M., \&
  {Schael}, A.~M. 2006{\natexlab{b}}, \mnras, 366, 609

\bibitem[{{Churchill} {et~al.}(1999){Churchill}, {Schneider}, {Schmidt}, \&
  {Gunn}}]{Churchill1999}
{Churchill}, C.~W., {Schneider}, D.~P., {Schmidt}, M., \& {Gunn}, J.~E. 1999,
  \aj, 117, 2573

\bibitem[{{Cicone} {et~al.}(2012){Cicone}, {Feruglio}, {Maiolino}, {Fiore},
  {Piconcelli}, {Menci}, {Aussel}, \& {Sturm}}]{Cicone2012}
{Cicone}, C., {Feruglio}, C., {Maiolino}, R., {et~al.} 2012, \aap, 543, A99

\bibitem[{{Cimatti} {et~al.}(2002){Cimatti}, {Pozzetti}, {Mignoli}, {Daddi},
  {Menci}, {Poli}, {Fontana}, {Renzini}, {Zamorani}, {Broadhurst}, {Cristiani},
  {D'Odorico}, {Giallongo}, \& {Gilmozzi}}]{Cimatti2002}
{Cimatti}, A., {Pozzetti}, L., {Mignoli}, M., {et~al.} 2002, \aap, 391, L1

\bibitem[{{Cirasuolo} {et~al.}(2010){Cirasuolo}, {McLure}, {Dunlop}, {Almaini},
  {Foucaud}, \& {Simpson}}]{Cirasuolo2010}
{Cirasuolo}, M., {McLure}, R.~J., {Dunlop}, J.~S., {et~al.} 2010, \mnras, 401,
  1166

\bibitem[{{Coil} {et~al.}(2011){Coil}, {Weiner}, {Holz}, {Cooper}, {Yan}, \&
  {Aird}}]{Coil2011}
{Coil}, A.~L., {Weiner}, B.~J., {Holz}, D.~E., {et~al.} 2011, \apj, 743, 46

\bibitem[{{Croton} {et~al.}(2006){Croton}, {Springel}, {White}, {De Lucia},
  {Frenk}, {Gao}, {Jenkins}, {Kauffmann}, {Navarro}, \& {Yoshida}}]{Croton2006}
{Croton}, D.~J., {Springel}, V., {White}, S.~D.~M., {et~al.} 2006, \mnras, 365,
  11

\bibitem[{{Cucciati} {et~al.}(2010){Cucciati}, {Iovino}, {Kova{\v c}},
  {Scodeggio}, {Lilly}, {Bolzonella}, {Bardelli}, {Vergani}, \&
  et~al.}]{Cucciati2010}
{Cucciati}, O., {Iovino}, A., {Kova{\v c}}, K., {et~al.} 2010, \aap, 524, A2

\bibitem[{{Daddi} {et~al.}(2005){Daddi}, {Renzini}, {Pirzkal}, {Cimatti},
  {Malhotra}, {Stiavelli}, {Xu}, {Pasquali}, {Rhoads}, {Brusa}, {di Serego
  Alighieri}, {Ferguson}, {Koekemoer}, {Moustakas}, {Panagia}, \&
  {Windhorst}}]{Daddi2005}
{Daddi}, E., {Renzini}, A., {Pirzkal}, N., {et~al.} 2005, \apj, 626, 680

\bibitem[{{Diamond-Stanic} {et~al.}(2012){Diamond-Stanic}, {Moustakas},
  {Tremonti}, {Coil}, {Hickox}, {Robaina}, {Rudnick}, \&
  {Sell}}]{Diamond-Stanic2012}
{Diamond-Stanic}, A.~M., {Moustakas}, J., {Tremonti}, C.~A., {et~al.} 2012,
  \apjl, 755, L26

\bibitem[{{Drory} {et~al.}(2003){Drory}, {Bender}, {Feulner}, {Hopp},
  {Maraston}, {Snigula}, \& {Hill}}]{Drory2003}
{Drory}, N., {Bender}, R., {Feulner}, G., {et~al.} 2003, \apj, 595, 698

\bibitem[{{Efstathiou}(2000)}]{Efstathiou2000}
{Efstathiou}, G. 2000, \mnras, 317, 697

\bibitem[{{Erb} {et~al.}(2006){Erb}, {Steidel}, {Shapley}, {Pettini}, {Reddy},
  \& {Adelberger}}]{Erb2006}
{Erb}, D.~K., {Steidel}, C.~C., {Shapley}, A.~E., {et~al.} 2006, \apj, 646, 107

\bibitem[{{Fazio} {et~al.}(2004){Fazio}, {Hora}, {Allen}, {Ashby}, {Barmby},
  {Deutsch}, {Huang}, {Kleiner}, {Marengo}, {Megeath}, {Melnick}, {Pahre},
  {Patten}, {Polizotti}, {Smith}, {Taylor}, {Wang}, {Willner}, \&
  et~al.}]{Fazio2004}
{Fazio}, G.~G., {Hora}, J.~L., {Allen}, L.~E., {et~al.} 2004, \apjs, 154, 10

\bibitem[{{Franx} {et~al.}(2003){Franx}, {Labb{\'e}}, {Rudnick}, {van Dokkum},
  {Daddi}, {F{\"o}rster Schreiber}, {Moorwood}, {Rix}, {R{\"o}ttgering}, {van
  der Wel}, {van der Werf}, \& {van Starkenburg}}]{Franx2003}
{Franx}, M., {Labb{\'e}}, I., {Rudnick}, G., {et~al.} 2003, \apjl, 587, L79

\bibitem[{{Furusawa} {et~al.}(2008){Furusawa}, {Kosugi}, {Akiyama}, {Takata},
  {Sekiguchi}, {Tanaka}, {Iwata}, {Kajisawa}, {Yasuda}, {Doi}, {Ouchi}, \&
  {Simpson}}]{Furusawa2008}
{Furusawa}, H., {Kosugi}, G., {Akiyama}, M., {et~al.} 2008, \apjs, 176, 1

\bibitem[{{Ganguly} {et~al.}(2013){Ganguly}, {Lynch}, {Charlton}, {Eracleous},
  {Tripp}, {Palma}, {Sembach}, {Misawa}, {Masiero}, {Milutinovic}, {Lackey}, \&
  {Jones}}]{Ganguly2013}
{Ganguly}, R., {Lynch}, R.~S., {Charlton}, J.~C., {et~al.} 2013, \mnras, 435,
  1233

\bibitem[{{Grogin} {et~al.}(2011){Grogin}, {Kocevski}, {Faber}, \&
  et~al.}]{Grogin2011}
{Grogin}, N.~A., {Kocevski}, D.~D., {Faber}, S.~M., \& et~al. 2011, \apjs, 197,
  35

\bibitem[{{Hainline} {et~al.}(2011){Hainline}, {Shapley}, {Greene}, \&
  {Steidel}}]{Hainline2011}
{Hainline}, K.~N., {Shapley}, A.~E., {Greene}, J.~E., \& {Steidel}, C.~C. 2011,
  \apj, 733, 31

\bibitem[{{Hall} {et~al.}(2002){Hall}, {Anderson}, {Strauss}, \&
  et~al.}]{Hall2002}
{Hall}, P.~B., {Anderson}, S.~F., {Strauss}, M.~A., \& et~al. 2002, \apjs, 141,
  267

\bibitem[{{Harrison} {et~al.}(2012){Harrison}, {Alexander}, {Swinbank},
  {Smail}, {Alaghband-Zadeh}, {Bauer}, {Chapman}, {Del Moro}, {Hickox},
  {Ivison}, {Men{\'e}ndez-Delmestre}, {Mullaney}, \& {Nesvadba}}]{Harrison2012}
{Harrison}, C.~M., {Alexander}, D.~M., {Swinbank}, A.~M., {et~al.} 2012,
  \mnras, 426, 1073

\bibitem[{{Hopkins} {et~al.}(2012){Hopkins}, {Quataert}, \&
  {Murray}}]{Hopkins2012b}
{Hopkins}, P.~F., {Quataert}, E., \& {Murray}, N. 2012, \mnras, 421, 3522

\bibitem[{{Kennicutt}(1998)}]{Kennicutt1998}
{Kennicutt}, Jr., R.~C. 1998, \araa, 36, 189

\bibitem[{{Koekemoer} {et~al.}(2011){Koekemoer}, {Faber}, {Ferguson}, \&
  et~al.}]{Koekemoer2011}
{Koekemoer}, A.~M., {Faber}, S.~M., {Ferguson}, H.~C., \& et~al. 2011, \apjs,
  197, 36

\bibitem[{{Kurk} {et~al.}(2013){Kurk}, {Cimatti}, {Daddi}, {Mignoli},
  {Pozzetti}, {Dickinson}, {Bolzonella}, {Zamorani}, {Cassata}, {Rodighiero},
  {Franceschini}, {Renzini}, {Rosati}, {Halliday}, \& {Berta}}]{Kurk2013}
{Kurk}, J., {Cimatti}, A., {Daddi}, E., {et~al.} 2013, \aap, 549, A63

\bibitem[{{Lagache} {et~al.}(2005){Lagache}, {Puget}, \& {Dole}}]{Lagache2005}
{Lagache}, G., {Puget}, J.-L., \& {Dole}, H. 2005, \araa, 43, 727

\bibitem[{{Lawrence} {et~al.}(2007){Lawrence}, {Warren}, {Almaini}, {Edge},
  {Hambly}, {Jameson}, {Lucas}, {Casali}, {Adamson}, {Dye}, {Emerson}, \&
  et~al.}]{Lawrence2007}
{Lawrence}, A., {Warren}, S.~J., {Almaini}, O., {et~al.} 2007, \mnras, 379,
  1599

\bibitem[{{Le F{\`e}vre} {et~al.}(2005){Le F{\`e}vre}, {Vettolani}, {Garilli},
  \& et~al.}]{LeFevre2005}
{Le F{\`e}vre}, O., {Vettolani}, G., {Garilli}, B., \& et~al. 2005, \aap, 439,
  845

\bibitem[{{Leitherer} {et~al.}(1999){Leitherer}, {Schaerer}, {Goldader},
  {Gonz{\'a}lez Delgado}, {Robert}, {Kune}, {de Mello}, {Devost}, \&
  {Heckman}}]{Leitherer1999}
{Leitherer}, C., {Schaerer}, D., {Goldader}, J.~D., {et~al.} 1999, \apjs, 123,
  3

\bibitem[{{Lilly} {et~al.}(2007){Lilly}, {Le F{\`e}vre}, {Renzini}, \&
  et~al.}]{Lilly2007}
{Lilly}, S.~J., {Le F{\`e}vre}, O., {Renzini}, A., \& et~al. 2007, \apjs, 172,
  70

\bibitem[{{Magnelli} {et~al.}(2011){Magnelli}, {Elbaz}, {Chary}, {Dickinson},
  {Le Borgne}, {Frayer}, \& {Willmer}}]{Magnelli2011}
{Magnelli}, B., {Elbaz}, D., {Chary}, R.~R., {et~al.} 2011, \aap, 528, A35

\bibitem[{{Martin}(2005)}]{Martin2005}
{Martin}, C.~L. 2005, \apj, 621, 227

\bibitem[{{Martin} {et~al.}(2012){Martin}, {Shapley}, {Coil}, {Kornei},
  {Bundy}, {Weiner}, {Noeske}, \& {Schiminovich}}]{Martin2012}
{Martin}, C.~L., {Shapley}, A.~E., {Coil}, A.~L., {et~al.} 2012, \apj, 760, 127

\bibitem[{{Misawa} {et~al.}(2007){Misawa}, {Charlton}, {Eracleous}, {Ganguly},
  {Tytler}, {Kirkman}, {Suzuki}, \& {Lubin}}]{Misawa2007}
{Misawa}, T., {Charlton}, J.~C., {Eracleous}, M., {et~al.} 2007, \apjs, 171, 1

\bibitem[{{Murphy} {et~al.}(2011){Murphy}, {Chary}, {Dickinson}, {Pope},
  {Frayer}, \& {Lin}}]{Murphy2011}
{Murphy}, E.~J., {Chary}, R.-R., {Dickinson}, M., {et~al.} 2011, \apj, 732, 126

\bibitem[{{Nesvadba} {et~al.}(2008){Nesvadba}, {Lehnert}, {De Breuck},
  {Gilbert}, \& {van Breugel}}]{Nesvadba2008}
{Nesvadba}, N.~P.~H., {Lehnert}, M.~D., {De Breuck}, C., {Gilbert}, A.~M., \&
  {van Breugel}, W. 2008, \aap, 491, 407

\bibitem[{{Papovich} {et~al.}(2006){Papovich}, {Moustakas}, {Dickinson}, {Le
  Floc'h}, {Rieke}, \& et~al.}]{Papovich2006}
{Papovich}, C., {Moustakas}, L.~A., {Dickinson}, M., {et~al.} 2006, \apj, 640,
  92

\bibitem[{{Pettini} {et~al.}(2002){Pettini}, {Rix}, {Steidel}, {Adelberger},
  {Hunt}, \& {Shapley}}]{Pettini2002}
{Pettini}, M., {Rix}, S.~A., {Steidel}, C.~C., {et~al.} 2002, \apj, 569, 742

\bibitem[{{Pozzetti} {et~al.}(2003){Pozzetti}, {Cimatti}, {Zamorani}, {Daddi},
  {Menci}, {Fontana}, {Renzini}, {Mignoli}, {Poli}, {Saracco}, {Broadhurst},
  {Cristiani}, {D'Odorico}, {Giallongo}, \& {Gilmozzi}}]{Pozzetti2003}
{Pozzetti}, L., {Cimatti}, A., {Zamorani}, G., {et~al.} 2003, \aap, 402, 837

\bibitem[{{Rieke} {et~al.}(2004){Rieke}, {Young}, {Engelbracht}, \&
  et~al.}]{Rieke2004}
{Rieke}, G.~H., {Young}, E.~T., {Engelbracht}, C.~W., \& et~al. 2004, \apjs,
  154, 25

\bibitem[{{Rodighiero} {et~al.}(2010){Rodighiero}, {Cimatti}, {Gruppioni},
  {Popesso}, {Andreani}, {Altieri}, {Aussel}, \& et~al.}]{Rodighiero2010}
{Rodighiero}, G., {Cimatti}, A., {Gruppioni}, C., {et~al.} 2010, \aap, 518, L25

\bibitem[{{Rupke} {et~al.}(2005{\natexlab{a}}){Rupke}, {Veilleux}, \&
  {Sanders}}]{Rupke2005a}
{Rupke}, D.~S., {Veilleux}, S., \& {Sanders}, D.~B. 2005{\natexlab{a}}, \apjs,
  160, 87

\bibitem[{{Rupke} {et~al.}(2005{\natexlab{b}}){Rupke}, {Veilleux}, \&
  {Sanders}}]{Rupke2005b}
{Rupke}, D.~S., {Veilleux}, S., \& {Sanders}, D.~B. 2005{\natexlab{b}}, \apjs,
  160, 115

\bibitem[{{Rupke} {et~al.}(2005{\natexlab{c}}){Rupke}, {Veilleux}, \&
  {Sanders}}]{Rupke2005c}
{Rupke}, D.~S., {Veilleux}, S., \& {Sanders}, D.~B. 2005{\natexlab{c}}, \apj,
  632, 751

\bibitem[{{Rupke} \& {Veilleux}(2013)}]{Rupke2013a}
{Rupke}, D.~S.~N. \& {Veilleux}, S. 2013, \apj, 768, 75

\bibitem[{{Sales} {et~al.}(2010){Sales}, {Navarro}, {Schaye}, {Dalla Vecchia},
  {Springel}, \& {Booth}}]{Sales2010}
{Sales}, L.~V., {Navarro}, J.~F., {Schaye}, J., {et~al.} 2010, \mnras, 409,
  1541

\bibitem[{{Saracco} {et~al.}(2005){Saracco}, {Longhetti}, {Severgnini}, {Della
  Ceca}, {Braito}, {Mannucci}, {Bender}, {Drory}, {Feulner}, {Hopp}, \&
  {Maraston}}]{Saracco2005}
{Saracco}, P., {Longhetti}, M., {Severgnini}, P., {et~al.} 2005, \mnras, 357,
  L40

\bibitem[{{Shapley} {et~al.}(2003){Shapley}, {Steidel}, {Pettini}, \&
  {Adelberger}}]{Shapley2003}
{Shapley}, A.~E., {Steidel}, C.~C., {Pettini}, M., \& {Adelberger}, K.~L. 2003,
  \apj, 588, 65

\bibitem[{{Smail} {et~al.}(2008){Smail}, {Sharp}, {Swinbank}, {Akiyama},
  {Ueda}, {Foucaud}, {Almaini}, \& {Croom}}]{Smail2008}
{Smail}, I., {Sharp}, R., {Swinbank}, A.~M., {et~al.} 2008, \mnras, 389, 407

\bibitem[{{Somerville} {et~al.}(2004){Somerville}, {Moustakas}, {Mobasher},
  {Gardner}, {Cimatti}, {Conselice}, {Daddi}, {Dahlen}, {Dickinson},
  {Eisenhardt}, {Lotz}, {Papovich}, {Renzini}, \& {Stern}}]{Somerville2004}
{Somerville}, R.~S., {Moustakas}, L.~A., {Mobasher}, B., {et~al.} 2004, \apjl,
  600, L135

\bibitem[{{Steidel} {et~al.}(2003){Steidel}, {Adelberger}, {Shapley},
  {Pettini}, {Dickinson}, \& {Giavalisco}}]{Steidel2003}
{Steidel}, C.~C., {Adelberger}, K.~L., {Shapley}, A.~E., {et~al.} 2003, \apj,
  592, 728

\bibitem[{{Steidel} {et~al.}(2010){Steidel}, {Erb}, {Shapley}, {Pettini},
  {Reddy}, {Bogosavljevi{\'c}}, {Rudie}, \& {Rakic}}]{Steidel2010}
{Steidel}, C.~C., {Erb}, D.~K., {Shapley}, A.~E., {et~al.} 2010, \apj, 717, 289

\bibitem[{{Sturm} {et~al.}(2011){Sturm}, {Gonz{\'a}lez-Alfonso}, {Veilleux},
  {Fischer}, {Graci{\'a}-Carpio}, {Hailey-Dunsheath}, {Contursi}, {Poglitsch},
  {Sternberg}, {Davies}, {Genzel}, {Lutz}, {Tacconi}, {Verma}, {Maiolino}, \&
  {de Jong}}]{Sturm2011}
{Sturm}, E., {Gonz{\'a}lez-Alfonso}, E., {Veilleux}, S., {et~al.} 2011, \apjl,
  733, L16

\bibitem[{{Talia} {et~al.}(2012){Talia}, {Mignoli}, {Cimatti}, {Kurk}, {Berta},
  {Bolzonella}, {Cassata}, {Daddi}, {Dickinson}, {Franceschini}, {Halliday},
  {Pozzetti}, {Renzini}, {Rodighiero}, {Rosati}, \& {Zamorani}}]{Talia2012}
{Talia}, M., {Mignoli}, M., {Cimatti}, A., {et~al.} 2012, \aap, 539, A61

\bibitem[{{Tasca} {et~al.}(2009){Tasca}, {Kneib}, {Iovino}, {Le F{\`e}vre},
  {Kova{\v c}}, {Bolzonella}, {Lilly}, \& et~al.}]{Tasca2009}
{Tasca}, L.~A.~M., {Kneib}, J.-P., {Iovino}, A., {et~al.} 2009, \aap, 503, 379

\bibitem[{{Tremonti} {et~al.}(2007){Tremonti}, {Moustakas}, \&
  {Diamond-Stanic}}]{Tremonti2007}
{Tremonti}, C.~A., {Moustakas}, J., \& {Diamond-Stanic}, A.~M. 2007, \apjl,
  663, L77

\bibitem[{{Trump} {et~al.}(2006){Trump}, {Hall}, {Reichard}, {Richards},
  {Schneider}, {Vanden Berk}, {Knapp}, {Anderson}, {Fan}, {Brinkman},
  {Kleinman}, \& {Nitta}}]{Trump2006}
{Trump}, J.~R., {Hall}, P.~B., {Reichard}, T.~A., {et~al.} 2006, \apjs, 165, 1

\bibitem[{{U} {et~al.}(2012){U}, {Sanders}, {Mazzarella}, {Evans}, {Howell},
  {Surace}, {Armus}, {Iwasawa}, {Kim}, {Casey}, {Vavilkin}, {Dufault},
  {Larson}, {Barnes}, {Chan}, {Frayer}, {Haan}, {Inami}, {Ishida},
  {Kartaltepe}, {Melbourne}, \& {Petric}}]{U2012}
{U}, V., {Sanders}, D.~B., {Mazzarella}, J.~M., {et~al.} 2012, \apjs, 203, 9

\bibitem[{{Ueda} {et~al.}(2008){Ueda}, {Watson}, {Stewart}, {Akiyama},
  {Schwope}, {Lamer}, {Ebrero}, {Carrera}, {Sekiguchi}, {Yamada}, {Simpson},
  {Hasinger}, \& {Mateos}}]{Ueda2008}
{Ueda}, Y., {Watson}, M.~G., {Stewart}, I.~M., {et~al.} 2008, \apjs, 179, 124

\bibitem[{{van Dokkum}(2001)}]{Dokkum2001}
{van Dokkum}, P.~G. 2001, \pasp, 113, 1420

\bibitem[{{Vanzella} {et~al.}(2009){Vanzella}, {Giavalisco}, {Dickinson},
  {Cristiani}, {Nonino}, {Kuntschner}, {Popesso}, {Rosati}, {Renzini}, {Stern},
  {Cesarsky}, {Ferguson}, \& {Fosbury}}]{Vanzella2009}
{Vanzella}, E., {Giavalisco}, M., {Dickinson}, M., {et~al.} 2009, \apj, 695,
  1163

\bibitem[{{Veilleux} {et~al.}(2013){Veilleux}, {Mel{\'e}ndez}, {Sturm},
  {Gracia-Carpio}, {Fischer}, {Gonz{\'a}lez-Alfonso}, {Contursi}, {Lutz},
  {Poglitsch}, {Davies}, {Genzel}, {Tacconi}, {de Jong}, {Sternberg}, {Netzer},
  {Hailey-Dunsheath}, {Verma}, {Rupke}, {Maiolino}, {Teng}, \&
  {Polisensky}}]{Veilleux2013}
{Veilleux}, S., {Mel{\'e}ndez}, M., {Sturm}, E., {et~al.} 2013, \apj, 776, 27

\bibitem[{{Verhamme} {et~al.}(2008){Verhamme}, {Schaerer}, {Atek}, \&
  {Tapken}}]{Verhamme2008}
{Verhamme}, A., {Schaerer}, D., {Atek}, H., \& {Tapken}, C. 2008, \aap, 491, 89

\bibitem[{{Weiner} {et~al.}(2009){Weiner}, {Coil}, {Prochaska}, {Newman},
  {Cooper}, {Bundy}, {Conselice}, {Dutton}, {Faber}, {Koo}, {Lotz}, {Rieke}, \&
  {Rubin}}]{Weiner2009}
{Weiner}, B.~J., {Coil}, A.~L., {Prochaska}, J.~X., {et~al.} 2009, \apj, 692,
  187

\bibitem[{{Weymann} {et~al.}(1991){Weymann}, {Morris}, {Foltz}, \&
  {Hewett}}]{Weymann1991}
{Weymann}, R.~J., {Morris}, S.~L., {Foltz}, C.~B., \& {Hewett}, P.~C. 1991,
  \apj, 373, 23

\end{thebibliography}

\label{lastpage}

\end{document}